\documentclass[prd,aps,showpacs,epsf,floats]{revtex4}
\usepackage{amsfonts}
\usepackage{amsmath}

\input{tcilatex}
\begin{document}

\title{Complexity Characterization in a Probabilistic Approach to Dynamical
Systems Through Information Geometry and Inductive Inference }
\author{S. A. Ali$^{1,2,3}$, C. Cafaro$^{4}$, A. Giffin$^{5}$, D.-H. Kim$^{1%
\text{, }6\text{, }7}$}
\affiliation{$^{1}$International Institute for Theoretical Physics and Mathematics
Einstein-Galilei, via Santa Gonda 14, 59100 Prato, Italy\\
$^{2}$Department of Physics, State University of New York at Albany, 1400
Washington Avenue, Albany, NY 12222, USA\\
$^{3}$Department of Arts and Sciences, Albany College of Pharmacy and Health
Sciences, 106 New Scotland Avenue, Albany, NY 12208, USA\\
$^{4}$School of Science and Technology, Physics Division, University of
Camerino, I-62032 Camerino, Italy\\
$^{5}$Princeton Institute for the Science and Technology of Materials,
Princeton University, Princeton, NJ 08540, USA\\
$^{6}$Center for Quantum Spacetime, Sogang University, Shinsu-dong 1,
Mapo-gu, Seoul 121-742, South Korea \\
$^{7}$Institute for the Early Universe, Ewha Womans University, Daehyun-dong
11-1, Seodaemun-gu, Seoul 120-750, South Korea}

\begin{abstract}
Information geometric techniques and inductive inference methods hold great
promise for solving computational problems of interest in classical and
quantum physics, especially with regard to complexity characterization of
dynamical systems in terms of their probabilistic description on curved
statistical manifolds. In this article, we investigate the possibility of
describing the macroscopic behavior of complex systems in terms of the
underlying statistical structure of their microscopic degrees of freedom by
use of statistical inductive inference and information geometry. We review
the Maximum Relative Entropy (MrE) formalism and the theoretical structure
of the information geometrodynamical approach to chaos (IGAC) on statistical
manifolds $\mathcal{M}_{S}$. Special focus is devoted to the description of
the roles played by the sectional curvature $\mathcal{K}_{\mathcal{M}_{S}}$,
the Jacobi field intensity $\mathcal{J}_{\mathcal{M}_{S}}$ and the
information geometrodynamical entropy $\mathcal{S}_{\mathcal{M}_{S}}$ (IGE).
These quantities serve as powerful information geometric complexity measures
of information-constrained dynamics associated with arbitrary chaotic and
regular systems defined on $\mathcal{M}_{S}$. Finally, the application of
such information geometric techniques to several theoretical models are
presented.
\end{abstract}

\maketitle

\section{Introduction}

It is commonly accepted that one of the major goals of physics is modeling
and predicting natural phenomena using relevant information about the system
of interest. Taking this statement seriously, it is reasonable to expect
that the laws of physics should reflect the methods for manipulating
information. Indeed, the less controversial opposite point of view may be
considered where the laws of physics are used to manipulate information.
This is exactly the point of view adopted in quantum information science
where information is manipulated using the laws of quantum mechanics \cite%
{nielsen}. In this work, we wish to explore an alternative viewpoint:
perhaps the laws of physics are nothing but rules of inference \cite%
{caticha-inference}. In this view the laws of physics are not laws of nature
but are merely the rules we follow when processing the information that
happens to be relevant to the physical problem under consideration. When the
information available is sufficient to make unequivocal, unique assessments
of truth we speak of making deductions: on the basis of this or that
information we deduce that a certain proposition is true. In cases where we
do not have statements that lead to unequivocal conclusions, we speak of
using inductive reasoning, and the system for this reasoning is probability
theory \cite{cox}. The word \textquotedblleft induction\textquotedblright\
refers to the process of using limited information about a few special cases
to draw conclusions about more general situations. Following this line of
reasoning, interesting probabilistic approaches describing complex dynamical
systems have already been investigated \cite{ca-7, ca-3, ca-5} .

The study of the relationship between entropy and the complexity \cite%
{gell-mann} of the trajectories of a dynamical system has always been an
active field of research \cite{brudno, blume, sze}. Chaotic behavior is a
particular case of complex behavior and it will be the subject of the
present work. In this article we make use of the so-called Entropic Dynamics
(ED) \cite{caticha1}. ED is a theoretical framework that arises from the
combination of inductive inference (Maximum Entropy Methods (ME), \cite%
{caticha2, sj, skilling, CG06, adom-tesi}) and Information Geometry (IG) 
\cite{amari_1}. The most intriguing question being pursued in ED stems from
the possibility of deriving dynamics from purely entropic arguments. This is
clearly valuable in circumstances where microscopic dynamics may be too far
removed from the phenomena of interest, such as in complex biological or
ecological systems, or where it may be unknown or perhaps even nonexistent,
as in economics. It has already been shown that entropic arguments do
account for a substantial part of the formalism of quantum mechanics, a
theory that is presumably fundamental \cite{caticha3}. Perhaps the
fundamental theories of physics are not so fundamental; they may just be
consistent, objective ways of manipulating information. Following this line
of thought, we extend the applicability of information geometric techniques
and inductive inference methods to computational problems of interest in
classical and quantum physics. In particular, we focus on the complexity
characterization of dynamical systems in terms of their probabilistic
description on curved statistical manifolds and identify relevant measures
of chaoticity associated with such an information geometrodynamical approach
to chaos (IGAC) \cite{ca-7, ca-3, ca-5, ca-1, ca-2, ca-4, ca-6}.

The layout of this article is as follows. In Section II, we review the
Maximum relative Entropy formalism. In Section III, we present an
introduction to the main features of the IGAC. In Section IV, we introduce
the information-geometric indicators of chaos for our theoretical \ model:
sectional $\left( \mathcal{K}_{\mathcal{M}_{S}}\right) $ and Ricci $(%
\mathcal{R}_{\mathcal{M}_{S}})$ curvatures, Jacobi field intensity $(%
\mathcal{J}_{\mathcal{M}_{S}})$, IGE $(\mathcal{S}_{\mathcal{M}_{S}})$ and
information geometric complexity $\left( \mathcal{C}_{\mathcal{M}%
_{S}}\right) $. In Section V, we present five applications of our techniques
to study the dynamical complexity of suitable statistical models. First, we
characterize the chaotic behavior of an ED Gaussian model describing an
arbitrary system of $l$ degrees of freedom in absence of correlations.
Second, we discuss the asymptotic temporal behavior of the information
geometric complexity of the maximum probability trajectories of a $2l$%
-dimensional Gaussian statistical manifold in presence of correlations
between the macrovariables labeling the macrostates of the system \cite{CM}.
Third, an information-geometric analogue of the Zurek-Paz quantum chaos
criterion of linear entropy growth \cite{ca-5, ca-2} for a system of
uncoupled, anisotropic (random frequency) and inverted Harmonic Oscillators
is examined. Forth example, we present the information-geometric
characterization of regular and chaotic quantum energy level statistics \cite%
{ca-7, ca-3}. Our fifth and final example applies the IGAC to model and
describe the scattering induced quantum entanglement of two
micro-correlated, spinless, structureless, non-relativistic particles, where
each particle is represented by a minimum uncertainty \textbf{wave-packet} 
\cite{PLA}. Final remarks are presented in Section VI.

\section{On the Maximum Relative Entropy Formalism}

In 1957, Jaynes \cite{jay} showed that maximizing statistical mechanic
entropy for the purpose of revealing how gas molecules were distributed was
equivalent to the maximizing of Shannon's information entropy \cite{shannon}
with statistical mechanical information. In traditional statistical
mechanics, Boltzmann applied relevant information regarding the gas
molecules in a closed system \emph{at equilibrium}, whereas Gibbs used
information relevant to an open system \emph{at equilibrium}. However, as
Jaynes noted, both solutions can be shown to be special cases of the maximum
entropy principle. It is important to emphasize that this method of
maximizing entropy is true for assigning probabilities regardless of the
information specifics \cite{adom-tesi}. As a consequence of this, the fact
that Boltzmann and Gibbs used information relevant at equilibrium does not
in general define entropy as only applying at equilibrium. Either of them
could have instead included information that was time dependent as well. It
is important to observe that Boltzmann included many assumptions into his
solutions, such as assuming that all microstates were independent and had
equal probability of occurring. The result of this was the interpretation
that his probability assignments were for individual microstates. Gibbs on
the other hand made no such assumptions. Therefore, his probability
assignments described the system as a whole. This makes Gibbs method more
general. In fact, it has been shown that Gibbs' entropy, not Boltzmann's, is
numerically equivalent to thermodynamic entropy \cite{jay1}.

Jaynes' use of the method of maximum entropy for assigning probabilities is
commonly known as \textit{MaxEnt} \cite{jay1}. This method has evolved to a
more general method, the method of Maximum (relative) Entropy (MrE) \cite%
{sj, skilling, CG06, adom-tesi} which has the advantage of not only
assigning probabilities but updating them (we use the term \textquotedblleft
updating\textquotedblright\ because as we gain valuable information, we
\textquotedblleft update\textquotedblright\ to new probabilities) when new
information is given in the form of constraints on the family of allowed
posteriors. This is similar in function to Bayes theorem, but using
macroscopic information (expectation values)\ as opposed to microscopic
information (data). We point out that one of the drawbacks of the MaxEnt
method was the inability to include data. Although we do not include
information in the form of data in this article, this problem has been
solved by showing that the MrE method can be used when either data,
constraint information or both is present \cite{GC07}. For our purposes
here, we will proceed to describe the simple case of how MrE is used to
update from one probability to another given macroscopic constraint
information only. Our first concern when using the MrE method to update from
a prior to a posterior distribution is to define the space in which the
search for the posterior will be conducted. Normally, when using the MrE
method, the updating is based on three pieces of information: prior
information (the prior), the known or assumed relationship between the sets
of microvariables and macrovariables (the model), and constraints on what
the new model or posterior is allowed to be. Some good general examples of
this are located in \cite{adom1, adom2}. However, for our purposes in the
present work we follow the simpler approach where our space of inquiry is
limited to one dimension of one or several quantities, $X\in \mathcal{X}$.
We intend to maximize an appropriate entropy form, 
\begin{equation}
\mathcal{S}[P,P_{\text{old}}]=-\int dX~P\left( X\right) \ln \frac{P\left(
X\right) }{P_{\text{old}}\left( X\right) },  \label{entropy}
\end{equation}%
subject to the appropriate constraints where $P(X)=P(x_{1},x_{2},\ldots
,x_{l})$, $l$ is the dimensionality of the microspace $\mathcal{X}$ and $P_{%
\text{old}}\left( X\right) $ contains our prior information. We impose the
usual normalization constraint, 
\begin{equation}
\int dX~P\left( X\right) =1,  \label{Normalization}
\end{equation}%
and include additional information about $X$ in the form of a constraint on
the expected value of some function $f\left( X\right) $, 
\begin{equation}
\int dX\,P\left( X\right) f\left( X\right) =\left\langle f\left( X\right)
\right\rangle =F.  \label{moment}
\end{equation}%
We proceed by maximizing (\ref{entropy}) subject to constraints (\ref%
{Normalization}) and (\ref{moment}). The purpose of maximizing the
logarithmic relative entropy $\mathcal{S}[P,P_{\text{old}}]$ is to determine
the value of $P\left( X\right) $ that is closest to $P_{\text{old}}\left(
X\right) $ given the normalization and information constraints. Using the
Lagrange multipliers formalism, we set the variation of\textbf{\ }$S[P%
\mathbf{,}P_{\text{old}}]$\textbf{\ }with respect to\textbf{\ }$P$\textbf{\ }%
equal to zero, to obtain%
\begin{equation}
\delta \left\{ \mathcal{S}[P,P_{\text{old}}]+\alpha \left[ \int dX\text{ }%
P\left( X\right) -1\right] +\beta \left[ \int dX\text{ }P\left( X\right)
f\left( X\right) -F\right] \right\} =0.  \label{mamma}
\end{equation}%
The quantities $\alpha $\textbf{\ }and\textbf{\ }$\beta $\textbf{\ }are
Lagrange multipliers whose actual values are determined by the value of the
constraints themselves. Substituting (\ref{entropy}) into (\ref{mamma}) we
get (after some algebra) 
\begin{equation}
-\int dX\left\{ \ln \frac{P\left( X\right) }{P_{\text{old}}\left( X\right) }%
+1-\alpha -\beta f\left( X\right) \right\} \delta P\left( X\right) =0,\text{ 
}\forall \delta P.  \label{mamma1}
\end{equation}%
Therefore, the terms inside the curly brackets in (\ref{mamma}) must sum to
zero, yielding 
\begin{equation}
P_{\text{new}}\left( X\right) =P_{\text{old}}\left( X\right) e^{\left[
-1+\alpha +\beta f\left( X\right) \right] }.  \label{Posterior 1}
\end{equation}%
In order to determine the Lagrange multipliers $\alpha $ and $\beta $, we
substitute our solution (\ref{Posterior 1}) into the constraint equations (%
\ref{Normalization}) and (\ref{moment}), respectively. Substituting (\ref%
{Posterior 1}) into (\ref{Normalization}), we obtain 
\begin{equation}
\int dX~P_{\text{old}}\left( X\right) e^{\beta f\left( X\right) }=e^{\left(
1-\alpha \right) }.  \label{mamma3}
\end{equation}%
Then substituting (\ref{mamma3}) into (\ref{Posterior 1}) yields 
\begin{equation}
P_{\text{new}}(X)=P_{\text{old}}\left( X\right) \frac{e^{\beta f\left(
X\right) }}{Z},  \label{Posterior 2}
\end{equation}%
where 
\begin{equation}
Z=e^{\left( 1-\alpha \right) }=\int dXe^{\beta f\left( X\right) }P_{\text{old%
}}\left( X\right) .
\end{equation}%
The Lagrange multiplier $\beta $ is determined by first substituting (\ref%
{Posterior 2}) into (\ref{moment}),%
\begin{equation}
\int dX\left[ P_{\text{old}}\left( X\right) \frac{e^{\beta f\left( X\right) }%
}{Z}\right] f\left( X\right) =F.  \label{F1}
\end{equation}%
Now, $\beta $ can be determined by rewriting (\ref{F1}) as 
\begin{equation}
\frac{\partial \ln Z}{\partial \beta }=F.  \label{F}
\end{equation}%
Finally, the selected posterior $P_{\text{new}}(X)$ has the expected
canonical form 
\begin{equation}
P_{\text{new}}(X)=P_{\text{old}}(X)\frac{e^{\beta f\left( X\right) }}{Z}.
\end{equation}%
When the expectation (\ref{moment}) is the first moment, the canonical form
is clearly seen. However, if the second moment is used a normal distribution
is produced, as in our example below. In conclusion, the ME method can be
described as a general method to update from a prior distribution $q(x)$ to
a posterior distribution $p(x)$ when new information becomes available. It
is perhaps worth emphasizing that in this approach, entropy is a tool for
reasoning which requires no interpretation in terms of heat, multiplicities,
disorder, uncertainty, or amount of information \cite{adom-tesi}.

In what follows we will use this inference methodology together with
information geometric methods to characterize the temporal complexity
(chaoticity, dynamical stochasticity) of suitable statistical models of
physical relevance.

\section{On the Information Geometry of Chaos}

IGAC\ is an application of ED to complex systems of arbitrary nature. It is
the information-geometric analogue of conventional geometrodynamical
approaches \cite{casetti_1, di bari_1, ca-1, ca-3, fox} where the classical
configuration space\textbf{\ }$\Gamma _{E}$\textbf{\ }is replaced by a
statistical manifold\textbf{\ }$\mathcal{M}_{S}$\textbf{\ }($\mathcal{M}_{S}$
is defined below in (\ref{stat-mani})) with the additional possibility of
considering chaotic dynamics arising from non conformally flat metrics (the
Jacobi metric is always conformally flat, instead). It is an
information-geometric extension of the Jacobi geometrodynamics (the
geometrization of a Hamiltonian system by transforming it to a geodesic flow 
\cite{jacobi_1}). The reformulation of dynamics in terms of a geodesic
problem allows for the application of a wide range of well-known geometrical
techniques to the investigation of the solution space and properties of the
equations of motion. The power of the Jacobi reformulation rests in the fact
that all dynamical information is collected into a single geometric object
(namely, the manifold on which geodesic flow is induced) in which all the
available manifest symmetries of the system are retained. For example, the
integrability of a system is connected with the existence of Killing vectors
and tensors on this manifold. The sensitive dependence of trajectories on
initial conditions, which is a key ingredient of chaos, can be investigated
from the equation of geodesic deviation. In the Riemannian \cite{casetti_1}
and Finslerian \cite{di bari_1} (a Finsler metric is obtained from a
Riemannian metric by relaxing the requirement that the metric be quadratic
on each tangent space) geometrodynamical approach to chaos in classical
Hamiltonian systems, an active field of research concerns the possibility%
\textbf{\ }of finding a rigorous relation among the sectional curvature, the
Lyapunov exponents, and the Kolmogorov-Sinai dynamical entropy (i.e., the
sum of positive Lyapunov exponents) \cite{kawabe_1}. Using
information-geometric methods, we have investigated in some detail the
aforementioned open research problem \cite{ca-1, ca-2, ca-3, ca-4, ca-5,
ca-6, ca-7}.

IGAC arises as a theoretical framework to study chaos in informational
geodesic flows describing physical, biological or chemical \ systems \cite%
{ca-3, ca-5, ca-7, fox}. The main goal of an ED model is that of inferring
\textquotedblleft macroscopic predictions\textquotedblright\ in the absence
of detailed knowledge of the microscopic nature of the arbitrary complex
systems being considered. More explicitly, by \textquotedblleft macroscopic
prediction\textquotedblright\ we mean knowledge of the statistical
parameters (expectation values) of the probability distribution function
that best reflects what is known about the system. This is an important
conceptual point. The probability distribution reflects the system in
general, not the microstates. We then select the relevant information about
the system. In other words, we have to select the macrospace of the system.
It is worth mentioning that the coexistence of macroscopic and microscopic
dynamics for a given physical (biological, chemical) system from a dynamical
and statistical point of view has always been a very important subject of
investigation \cite{kaneto}. For example, in certain fluid systems showing
Rayleigh-Benard convection \cite{lorenz}, the macroscopic chaotic behavior
(macroscopic chaos) is a manifestation of the underlying molecular
interaction of a very large collection of molecules (microscopic or
molecular chaos). Moreover, macroscopic chaos arising from an underlying
microscopic chaotic molecular behavior has been observed in chemical
reactions. In order to study the underlying dynamics in rate equations of
chemical reactions, a mesoscopic description has been adopted which is given
by a set of transition probabilities among chemicals. In such a description,
the underlying dynamics of macroscopic motion is that of stochastic
processes and the evolution of the probability distribution associated with
each chemical is investigated \cite{fox}.

In what follows, we schematically outline the main features underlying the
construction of an arbitrary form of entropic dynamics. First, the
microstates of the system under investigation must be defined. For the sake
of simplicity, we assume the system is characterized by an\textbf{\ }$l$%
-dimensional microspace $\mathcal{X}$ with microstates $X\equiv \left(
x_{1},\ldots ,x_{l}\right) $. The quantities $x_{k}$ with $k=1,2,\ldots ,l$
denote the degrees of freedom of the microstates\textbf{\ }of the system. We
assume that each degree of freedom $x_{k}$ is subject to $n$-information
constraints. The $n$-information constraints may (though not necessarily)
correspond to the moments of $P\left( X|\Theta \right) $. For example, if we
consider a system in which the relevant information constrains are given by
the first two moments of $P\left( X|\Theta \right) $, namely the expectation
values $\mu _{k}$ and variances $\sigma _{k}$,%
\begin{equation}
\left\langle x_{k}\right\rangle \equiv \mu _{k}\text{ and }\left(
\left\langle \left( x_{k}-\left\langle x_{k}\right\rangle \right)
^{2}\right\rangle \right) ^{\frac{1}{2}}\equiv \sigma _{k},  \label{var}
\end{equation}%
then the resulting posterior probability distribution function is a Gaussian
distribution $p_{k}\left( x_{k}|\mu _{k}\text{, }\sigma _{k}\right) $. For
general purposes, it is convenient to define an $nl$-dimensional macroscopic
vector $\Theta $ with statistical coordinates $\left\{ \vartheta
_{k}^{\left( m\right) }\right\} $ (where $k=1,2,\ldots ,l$ labels the
microstates and $m=1,2,\ldots ,n$ enumerates the information constraints).
The set $\Theta $ defines the $nl$-dimensional space of macrostates of the
system, namely the parameter space $\mathcal{I}_{\Theta }$. In the \ $l$%
-dimensional Gaussian case, the vector $\Theta $ is $nl$-dimensional and has
coordinates $\left( \mu _{1},\ldots ,\mu _{l};\sigma _{1},\ldots ,\sigma
_{l}\right) $.

At this point we make a working hypothesis that the microstates are
statistically independent. Then, in addition to information constraints, each%
\textbf{\ }distribution\textbf{\ }$p_{k}\left( x_{k}|\vartheta _{k}^{\left(
m\right) }\right) $\textbf{\ }of each degree of freedom $x_{k}$\ must
satisfy the usual normalization conditions%
\begin{equation}
\dint\limits_{\left\{ x_{k}\right\} }dx_{k}p_{k}\left( x_{k}|\vartheta
_{k}^{\left( m\right) }\right) =1.
\end{equation}%
Once the microstates have been defined and the relevant (linear or
nonlinear) information constraints selected, we are left with a set of $l$%
-dimensional vector probability distributions 
\begin{equation}
P\left( X|\Theta \right) =\underset{k=1}{\overset{l}{\dprod }}p_{k}\left(
x_{k}|\vartheta _{k}^{\left( m\right) }\right)  \label{PDG}
\end{equation}%
(assuming statistical independence, identically distributed microvariables
and assuming as an additional working hypothesis that the prior probability
distribution is uniform) encoding the relevant available information about
the system.

The statistical manifold $\mathcal{M}_{S}$,%
\begin{equation}
\mathcal{M}_{S}=\left\{ P\left( X|\Theta \right) =\underset{k=1}{\overset{l}{%
\dprod }}p_{k}\left( x_{k}|\vartheta _{k}^{\left( m\right) }\right) \right\}
\label{stat-mani}
\end{equation}%
is defined as the set of probabilities $\left\{ P\left( X|\Theta \right)
\right\} $ described above with $X\in 
\mathbb{R}
^{l}$, $\Theta \in \mathcal{I}_{\Theta }$ where $\mathcal{I}_{\Theta }$ is
comprised of the $n$ parameter sub-spaces corresponding to each of the $n$%
-information constraint according to $\mathcal{I}_{\Theta
}=\bigotimes\limits_{m=1}^{n}\mathcal{I}^{(m)}\left( \vartheta _{k}^{\left(
m\right) }\right) $. Each point of the geodesic on an $nl$-dimensional
statistical manifold $\mathcal{M}_{S}$ represents a macrostate $\Theta $
parametrized by the macroscopic dynamical variables $\left\{ \vartheta
_{k}^{\left( m\right) }\right\} $ defining the macrostate of the system.
Furthermore, each macrostate is in one-to-one correspondence with the
probability distribution $P\left( X|\Theta \right) $ representing the
maximally probable description of the system being considered. Thus, the set
of macrostates forms the parameter space $\mathcal{I}_{\Theta }$ while the
set of probability distributions forms the statistical manifold $\mathcal{M}%
_{S}$. Referring to our Gaussian example above, $\mathcal{I}_{\Theta }$ is
the direct product of the parameter sub-spaces $\mathcal{I}_{\mu }$
(corresponding to the first moment $n=1$ of the Gaussian, i.e., the
expectation $\mu $) and $\mathcal{I}_{\sigma }$ (corresponding to the second
moment $n=2$ of the Gaussian, i.e., the variance $\sigma $), where $\mathcal{%
I}_{\mu }=\left( -\infty ,+\infty \right) _{\mu }$ and $\mathcal{I}_{\sigma
}=\left( 0,+\infty \right) _{\sigma }$.

A measure of distinguishability among macrostates is obtained by assigning a
probability distribution $P\left( X|\Theta \right) \ni \mathcal{M}_{S}$ to
each macrostate $\Theta $. Assignment of a probability distribution to each
state endows $\mathcal{M}_{S}$ with a metric structure. Specifically, the
Fisher-Rao information metric $g_{ab}\left( \Theta \right) $ \cite{amari_1}, 
\begin{equation}
g_{ab}\left( \Theta \right) =\int dXP\left( X|\Theta \right) \partial
_{a}\ln P\left( X|\Theta \right) \partial _{b}\ln P\left( X|\Theta \right) ;%
\text{ }a,b=1,\ldots ,nl\text{ and }\partial _{a}=\frac{\partial }{\partial
\vartheta ^{a}},  \label{fisher-rao}
\end{equation}%
defines a measure of distinguishability among macrostates on $\mathcal{M}%
_{S} $.

It is known from IG \cite{amari_1} that there is a one-to-one relation
between elements of the statistical manifold $\mathcal{M}_{s}$ and the
parameter space $\mathcal{I}_{\Theta }$. Specifically, the statistical
manifold $\mathcal{M}_{s}$ is homeomorphic to the parameter space $\mathcal{I%
}_{\Theta }$. This implies the existence of a continuous, bijective map $h_{%
\mathcal{M}_{s},\mathcal{I}_{\Theta }}$,%
\begin{equation}
h_{\mathcal{M}_{s},\mathcal{I}_{\Theta }\text{ }}:\mathcal{M}_{S}\ni P\left(
X|\Theta \right) \rightarrow \Theta \in \mathcal{I}_{\Theta },
\end{equation}%
where $h_{\mathcal{M}_{s},\mathcal{I}_{\Theta }\text{ }}^{-1}\left( \Theta
\right) =P\left( X|\Theta \right) $. The inverse image $h_{\mathcal{M}_{s},%
\mathcal{I}_{\Theta }\text{ }}^{-1}$ is the so-called homeomorphism map. It
is worth pointing out that the possible chaotic behavior of the set of
macrostates\textbf{\ }$\Theta $\textbf{\ }is strictly related to the
selected relevant information about the set of microstates\textbf{\ }$X$%
\textbf{\ }of the system. In other words, the assumed Gaussian
characterization of the degrees of freedom\textbf{\ }$\left\{ x_{k}\right\} $%
\ of each microstate of the system has deep consequences on the macroscopic
behavior of the system itself. More generally, within our theoretical
construct, \emph{\textquotedblleft the macroscopic behavior of an arbitrary
complex system is a consequence of the underlying statistical structure of
the microscopic degrees of freedom of the system being
considered\textquotedblright }.

It should be noted that coupled constraints would lead to a
\textquotedblleft generalized\textquotedblright\ product rule in (\ref{PDG})
and to a metric tensor (\ref{fisher-rao}) with non-trivial off-diagonal
elements (covariance terms). In presence of correlated degrees of freedom $%
\left\{ x_{j}\right\} $, the \textquotedblleft
generalized\textquotedblright\ product rule becomes%
\begin{equation}
P_{\text{tot}}\left( x_{1},\ldots ,x_{n}\right)
=\dprod\limits_{j=1}^{n}P_{j}\left( x_{j}\right) \overset{\text{correlations}%
}{\longrightarrow }P_{\text{tot}}^{\prime }\left( x_{1},\ldots ,x_{n}\right)
\neq \dprod\limits_{j=1}^{n}P_{j}\left( x_{j}\right) ,
\end{equation}%
where%
\begin{equation}
P_{\text{tot}}^{\prime }\left( x_{1},\ldots ,x_{n}\right) \overset{\text{def}%
}{=}P_{n}\left( x_{n}|x_{1},\ldots ,x_{n-1}\right) P_{n-1}\left(
x_{n-1}|x_{1},\ldots ,x_{n-2}\right) \cdots P_{2}\left( x_{2}|x_{1}\right)
P_{1}\left( x_{1}\right) .
\end{equation}%
For instance, correlations in the degrees of freedom may be introduced in
terms of the following information-constraints,%
\begin{equation}
x_{j}=f_{j}\left( x_{1},\ldots ,x_{j-1}\right) ,\text{ }\forall j=2,\ldots
,n.
\end{equation}%
In such a case, we obtain%
\begin{equation}
P_{\text{tot}}^{\prime }\left( x_{1},\ldots ,x_{n}\right) =\delta \left(
x_{n}-f_{n}\left( x_{1},\ldots ,x_{n-1}\right) \right) \delta \left(
x_{n-1}-f_{n-1}\left( x_{1},\ldots ,x_{n-2}\right) \right) \cdots \delta
\left( x_{2}-f_{2}\left( x_{1}\right) \right) P_{1}\left( x_{1}\right) ,
\end{equation}%
where the $j$-th probability distribution $P_{j}\left( x_{j}\right) $ is
given by%
\begin{equation}
P_{j}\left( x_{j}\right) =\int \cdots \int dx_{1}\cdots
dx_{j-1}dx_{j+1}\cdots dx_{n}P_{\text{tot}}^{\prime }\left( x_{1},\ldots
,x_{n}\right) .
\end{equation}%
Correlations between the microscopic degrees of freedom of the system $%
\left\{ x_{j}\right\} $ (micro-correlations) are conventionally introduced
by means of the correlation coefficients $r_{ij}^{\left( \text{micro}\right)
}$, 
\begin{equation}
r_{ij}^{\left( \text{micro}\right) }=r\left( x_{i},x_{j}\right) \overset{%
\text{def}}{=}\frac{\left\langle x_{i}x_{j}\right\rangle -\left\langle
x_{i}\right\rangle \left\langle x_{j}\right\rangle }{\sigma _{i}\sigma _{j}};%
\text{ }\sigma _{i}=\sqrt{\left\langle \left( x_{i}-\left\langle
x_{i}\right\rangle \right) ^{2}\right\rangle },  \label{microC}
\end{equation}%
with $r_{ij}^{\left( \text{micro}\right) }\in \left( -1,1\right) $ and $%
i,j=1,\ldots ,n$. For the $2n$-dimensional Gaussian statistical model in
presence of micro-correlations, the system is described by the following
probability distribution $P\left( X|\Theta \right) $,%
\begin{equation}
P\left( X|\Theta \right) =\frac{1}{\left[ \left( 2\pi \right) ^{n}\det
C\left( \Theta \right) \right] ^{\frac{1}{2}}}\exp \left[ -\frac{1}{2}\left(
X-M\right) ^{t}\cdot C^{-1}\left( \Theta \right) \cdot \left( X-M\right) %
\right] \neq \dprod\limits_{j=1}^{n}\left( 2\pi \sigma _{j}^{2}\right) ^{-%
\frac{1}{2}}\exp \left[ -\frac{\left( x_{j}-\mu _{j}\right) ^{2}}{2\sigma
_{j}^{2}}\right] ,  \label{CG}
\end{equation}%
where $X=\left( x_{1},\ldots ,x_{n}\right) $, $M=\left( \mu _{1},\ldots ,\mu
_{n}\right) $ and $C\left( \Theta \right) $ is the $\left( 2n\times
2n\right) $-dimensional (non-singular) covariance matrix.

Once $\mathcal{M}_{S}$ and $\mathcal{I}_{\Theta }$ are defined, the ED
formalism provides the tools to explore dynamics driven \ on $\mathcal{M}%
_{S} $\ by entropic arguments. Specifically, given a known initial
macrostate $\Theta ^{\left( \text{initial}\right) }$ (probability
distribution), and that the system evolves to a final known macrostate $%
\Theta ^{\left( \text{final}\right) }$, the possible trajectories of the
system are examined in the ED approach using ME methods. We emphasize ED can
be derived from a standard principle of least action
(Maupertuis-Euler-Lagrange-Jacobi-type) \cite{caticha1, arnold}. The main
differences are that the dynamics being considered here, namely Entropic
Dynamics, is defined on a space of probability distributions $\mathcal{M}%
_{s} $, not on an ordinary linear space $V$ and the standard coordinates $%
q_{a}$ of the system are replaced by statistical macrovariables $\vartheta
^{a}$. The geodesic equations for the macrovariables of the Gaussian ED
model are given by\textit{\ }nonlinear second order coupled ordinary
differential equations%
\begin{equation}
\frac{d^{2}\vartheta ^{a}}{d\tau ^{2}}+\Gamma _{bc}^{a}\frac{d\vartheta ^{b}%
}{d\tau }\frac{d\vartheta ^{c}}{d\tau }=0.  \label{geodesic equations}
\end{equation}%
The geodesic equations in (\ref{geodesic equations}) describe reversible
dynamics whose solution is the trajectory between an initial $\Theta
^{\left( \text{initial}\right) }$ and a final macrostate $\Theta ^{\left( 
\text{final}\right) }$. The trajectory can be traversed equally well in both
directions. A geodesic on a curved statistical manifold $\mathcal{M}_{S}$
represents the maximum probability path a complex dynamical system explores
in its evolution between initial and final macrostates $\Theta ^{\left( 
\text{initial}\right) }$ and $\Theta ^{\left( \text{final}\right) }$,
respectively. We point out that this path is obtained by use of a Principle
of Probable Inference, namely the Maximum relative Entropy Method. Our
theoretical formalism allows us to analyze important physics problems
through statistical inference and information geometric techniques.

\section{Information Geometric Measures of Temporal Complexity}

In this Section, mainly following one of our previous works \cite{ca-6}, we
introduce the relevant indicators of chaoticity (temporal complexity,
dynamical stochasticity; in general there is no one-to-one relation between
chaos and complexity, chaos may imply complexity but necessarily vice versa)
within our theoretical formalism. Specifically, once the Fisher-Rao
information metric is given, we apply standard methods of Riemannian
differential geometry to study the information-geometric structure of the
manifold $\mathcal{M}_{S}$ underlying the entropic dynamics. Connection
coefficients $\Gamma _{bc}^{a}$, Ricci tensor $\mathcal{R}_{ab}$, Riemannian
curvature tensor $\mathcal{R}_{abcd}$, sectional curvatures $\mathcal{K}_{%
\mathcal{M}_{S}}$ \cite{MTW}, scalar curvature $\mathcal{R}_{\mathcal{M}%
_{S}} $, Weyl anisotropy tensor $\mathcal{W}_{abcd}$ (the anisotropy of the
manifold underlying the system dynamics plays a significant role in the
mechanism of instability), the Jacobi vector field intensity $\mathcal{J}_{%
\mathcal{M}_{S}}$, the information-geometric analogue of the Lyapunov
exponent $\lambda _{\mathcal{J}}$ and the information-geometric entropy
(IGE) $\mathcal{S}_{\mathcal{M}_{s}}$ can be calculated in the standard
manner \cite{casetti_1, di bari_1, ca-6}.

In order to characterize the chaotic behavior of complex entropic dynamical
systems, we are primarily concerned with the signs of the scalar and
sectional curvatures $\mathcal{K}_{\mathcal{M}_{S}}$ of $\mathcal{M}_{S}$,
the asymptotic behavior of Jacobi fields $J^{a}$ on $\mathcal{M}_{S}$, the
existence a non-vanishing Weyl anisotropy tensor $\mathcal{W}_{abcd}$ and
the asymptotic behavior of the IGE $\mathcal{S}_{\mathcal{M}_{S}}$. It is
crucial to observe that true chaos is identified by the occurrence of two
features \cite{di bari_1}: 1) strong dependence on initial conditions and
exponential divergence of the Jacobi vector field intensity, i.e. stretching
of dynamical trajectories; 2) compactness of the configuration space
manifold, i.e. folding of dynamical trajectories.

\subsection{Sectional and Scalar Curvatures}

Once the Fisher-Rao information metric\textbf{\ }$g_{ab}$\textbf{\ }is given%
\textbf{,} we use standard differential geometry methods applied to the
space of probability distributions to characterize the geometric properties
of $\mathcal{M}_{s}$. Recall that the Ricci scalar curvature $\mathcal{R}$
is given by%
\begin{equation}
\mathcal{R}=g^{ab}\mathcal{R}_{ab},  \label{ricci-scalar}
\end{equation}%
where $g^{ab}g_{bc}=\delta _{\text{ }c}^{a}$ so that $g^{ab}=\left(
g_{ab}\right) ^{-1}$. The Ricci tensor $\mathcal{R}_{ab}$ is given by%
\begin{equation}
\mathcal{R}_{ab}=\partial _{c}\Gamma _{ab}^{c}-\partial _{b}\Gamma
_{ac}^{c}+\Gamma _{ab}^{c}\Gamma _{cd}^{d}-\Gamma _{ac}^{d}\Gamma _{bd}^{c}.
\label{ricci-tensor}
\end{equation}%
The Christoffel symbols $\Gamma _{ab}^{c}$ appearing in the Ricci tensor are
defined in the standard manner as 
\begin{equation}
\Gamma _{ab}^{c}=\frac{1}{2}g^{cd}\left( \partial _{a}g_{db}+\partial
_{b}g_{ad}-\partial _{d}g_{ab}\right) .  \label{connection1}
\end{equation}%
It can be shown that the Ricci scalar curvature can be written as \cite{MTW}%
\begin{equation}
\mathcal{R}_{\mathcal{M}_{s}}=\mathcal{R}_{abcd}g^{ac}g^{bd}=\mathcal{R}_{%
\text{ }a}^{a}=\sum_{i\neq j}\mathcal{K}\left( e_{i},e_{j}\right) .
\label{Ricci}
\end{equation}%
The scalar curvature is the sum of all sectional curvatures $\mathcal{K}%
\left( e_{i},e_{j}\right) $ of planes spanned by pairs of orthonormal basis
elements $\left\{ e_{a}=\partial _{\vartheta ^{a}(p)}\right\} $ of the
tangent space $T_{p}\mathcal{M}_{s}$ with $p\in \mathcal{M}_{s}$ \cite{MTW}, 
\begin{equation}
\mathcal{K}\left( u,v\right) =\frac{\mathcal{R}_{abcd}u^{a}v^{b}u^{c}v^{d}}{%
\left( g_{ad}g_{bc}-g_{ac}g_{bd}\right) u^{a}v^{b}u^{c}v^{d}};\text{ }%
u\rightarrow h^{i},\text{ }v\rightarrow h^{j}\text{ with }i\neq j,
\label{sectionK}
\end{equation}%
where $\left\langle e_{a},h^{b}\right\rangle =\delta _{\text{ }a}^{b}$.
Notice that the sectional curvatures completely determine the curvature
tensor.

The negativity of the Ricci scalar $\mathcal{R}_{\mathcal{M}_{S}}$ implies
the existence of expanding directions in the configuration space manifold $%
\mathcal{M}_{s}$. Indeed, since $\mathcal{R}_{\mathcal{M}_{S}}$ is the sum
of all sectional curvatures of planes spanned by pairs of orthonormal basis
elements $\left\{ e_{a}=\partial _{\vartheta ^{a}}\right\} $, the negativity
of the Ricci scalar is only a sufficient (not necessary) condition for local
instability of geodesic flow. For this reason, the negativity of the scalar
provides a strong\textit{\ }criterion of local instability. Scenarios may
arise where negative sectional curvatures are present, but the positive ones
could prevail in the sum so that the Ricci scalar is non-negative despite
the instability in the flow in those directions. Consequently, the signs of $%
\mathcal{K}_{\mathcal{M}_{S}}$ are of crucial significance for the proper
characterization of chaos.

\subsection{Killing Vectors}

Yet another useful way to understand the anisotropy of the $\mathcal{M}_{s}$
is the following. It is known that in $N$ dimensions, there are at most $%
\frac{N\left( N+1\right) }{2}$ independent Killing vectors (directions of
symmetry of the manifold). Since $\mathcal{M}_{s}$ is not a pseudosphere,
the information metric tensor does not admit the maximum number of Killing
vectors $K_{a}$ defined as%
\begin{equation}
\mathcal{L}_{K}g_{ab}=\mathcal{D}_{a}K_{b}+\mathcal{D}_{b}K_{a}=0,
\end{equation}%
where $\mathcal{D}_{a}$, defined as%
\begin{equation}
\mathcal{D}_{a}K_{b}=\partial _{a}K_{b}-\Gamma _{ba}^{c}K_{c},
\end{equation}%
is the covariant derivative operator with respect to the connection $\Gamma $
defined in (\ref{connection1}). The Lie derivative $\mathcal{L}_{K}g_{ab}$
of the tensor field $g_{ab}$ along a given direction $K$ measures the
intrinsic variation of the field along that direction (that is, the metric
tensor is Lie transported along the Killing vector) \cite{clarke}. Locally,
a maximally symmetric space of Euclidean signature is either a plane, a
sphere, or a hyperboloid, depending on the sign of $\mathcal{R}$. In our
case, none of these scenarios occur. As will be seen in what follows, this
fact has a significant impact on the integration of the geodesic deviation
equation on $\mathcal{M}_{s}$. At this juncture, we emphasize it is known
that the anisotropy of the manifold underlying system dynamics plays a
crucial role in the mechanism of instability. In particular, fluctuating
sectional curvatures require also that the manifold be anisotropic. However,
the connection between curvature variations along geodesics and anisotropy
is far from clear \cite{ca-6}. Krylov was the first to emphasize \cite%
{krylov} the use of $\mathcal{R}<0$ as an instability criterion in the
context of an $N$-body system (a gas) interacting via Van der Waals forces,
with the ultimate hope to understand the relaxation process in a gas.
However, Krylov neglected the problem of compactness of the configuration
space manifold which is important for making inferences about exponential
mixing of geodesic flows \cite{pellicott}. Compactness \cite{jost, cipriani}
is required in order to discard trivial exponential growths due to the
unboundedness of the \textquotedblleft volume\textquotedblright\ available
to the dynamical system. In other words, the folding is necessary to have a
dynamics that is actually able to mix the trajectories, making practically
impossible, after a finite interval of time, to discriminate between
trajectories which were very nearby each other at the initial time. When the
space is not compact, even in presence of strong dependence on initial
conditions, it could be possible in some instances (though not always), to
distinguish among different trajectories originating within a small distance
and then evolved subject to exponential instability. As a final remark, we
note that since homeomorphisms preserve compactness and since $\mathcal{M}%
_{s}$ and $\mathcal{I}_{\Theta }$ are homeomorphic, it is sufficient to show
that the parameter space $\mathcal{I}_{\Theta }$ is compact in order to
ensure that the corresponding submanifold of $\mathcal{M}_{s}$ is itself
compact.

\subsection{Jacobi Fields}

A powerful mathematical tool to investigate the stability or instability of
a geodesic flow is the Jacobi-Levi-Civita equation (JLC equation) for
geodesic spread \cite{casetti_1}. The JLC-equation covariantly describes how
nearby geodesics locally scatter and relates the stability or instability of
a geodesic flow with curvature properties of the ambient manifold. For the
sake of clarity, consider the behavior of a family of neighboring geodesics $%
\left\{ \vartheta _{\mathcal{M}_{s}}^{a}\left( \tau ;\vec{\varsigma}\right)
\right\} _{\vec{\varsigma}\in \mathbb{R}^{+N}}^{a=1,\ldots ,N}$ on a
statistical manifold $\mathcal{M}_{s}$ with dim$_{%
\mathbb{R}
}\mathcal{M}_{s}=N$. The geodesics $\vartheta _{\mathcal{M}_{s}}^{a}\left(
\tau ;\vec{\varsigma}\right) $ are solutions of equation (\ref{geodesic
equations}). The relative geodesic spread on a non-maximally symmetric
curved manifold as $\mathcal{M}_{s}$ is characterized by the
Jacobi-Levi-Civita (JLC) equation \cite{MTW, carmo}%
\begin{equation}
\frac{\mathcal{D}^{2}J^{a}}{\mathcal{D}\tau ^{2}}+\mathcal{R}_{\text{ }%
bcd}^{a}\frac{\partial \vartheta ^{b}}{\partial \tau }J^{c}\frac{\partial
\vartheta ^{d}}{\partial \tau }=0,  \label{JLC-1}
\end{equation}%
where the covariant derivative $\frac{\mathcal{D}^{2}J^{a}}{\mathcal{D}\tau
^{2}}$\textbf{\ }in (\ref{JLC-1}) is defined as \cite{ohanian}%
\begin{eqnarray}
\frac{\mathcal{D}^{2}J^{a}}{\mathcal{D}\tau ^{2}} &=&\frac{d^{2}J^{a}}{d\tau
^{2}}+2\Gamma _{bc}^{a}\frac{dJ^{b}}{d\tau }\frac{d\vartheta ^{c}}{d\tau }%
+\Gamma _{bc}^{a}J^{b}\frac{d^{2}\vartheta ^{c}}{d\tau ^{2}}+\Gamma
_{bc,d}^{a}\frac{d\vartheta ^{d}}{d\tau }\frac{d\vartheta ^{c}}{d\tau }J^{b}+
\notag \\
&&+\Gamma _{bc}^{a}\Gamma _{df}^{b}\frac{d\vartheta ^{f}}{d\tau }\frac{%
d\vartheta ^{c}}{d\tau }J^{d},
\end{eqnarray}%
and the Jacobi vector field $J^{a}$ is given by \cite{defelice}%
\begin{equation}
J^{a}=\delta _{\vec{\varsigma}}\vartheta ^{a}\equiv \left. \frac{\partial
\vartheta ^{a}\left( \tau ;\vec{\varsigma}\right) }{\partial \varsigma ^{b}}%
\right\vert _{\tau }\delta \varsigma ^{b}.  \label{J1}
\end{equation}%
Equation (\ref{JLC-1}) forms a system of $N$ coupled ordinary differential
equations linear in the components of the deviation vector field (\ref{J1})
but\textit{\ }nonlinear in derivatives of the metric (\ref{fisher-rao}). It
describes the linearized geodesic flow: the linearization ignores the
relative velocity of the geodesics. When the geodesics are neighboring but
their relative velocity is arbitrary, the corresponding geodesic deviation
equation is the so-called generalized Jacobi equation \cite{chicone}. The
nonlinearity is due to the existence of velocity-dependent terms in the
system. Neighboring geodesics accelerate relative to each other with a rate
directly measured by the curvature tensor $\mathcal{R}_{abcd}$. The
non-trivial integration of (\ref{JLC-1}) leads to the following expression
of the Jacobi vector field intensity,%
\begin{equation}
\mathcal{J}_{\mathcal{M}_{S}}\equiv \left\Vert J^{a}\right\Vert \overset{%
\text{def}}{=}\left( g_{ab}J^{a}J^{b}\right) ^{\frac{1}{2}}.
\label{eq:VII-9}
\end{equation}%
For applications of the asymptotic temporal behavior of $\mathcal{J}_{%
\mathcal{M}_{S}}\left( \tau \right) $ as a reliable indicator of chaoticity,
we refer to our previous articles in references \cite{ca-1, ca-2, ca-3}. The
geodesic spread on $\mathcal{M}_{s}$ is described by means of an
exponentially divergent Jacobi vector field intensity$\mathcal{J}_{\mathcal{M%
}_{s}}$, a classical feature of chaos. In our approach, inspired by the work
presented in \cite{PR}, the quantity $\lambda _{\mathcal{M}_{S}}$ defined as%
\begin{equation}
\lambda _{\mathcal{M}_{S}}\overset{\text{def}}{=}\lim_{\tau \rightarrow
\infty }\left[ \frac{1}{\tau }\ln \left( \frac{\left\vert \mathcal{J}_{%
\mathcal{M}_{S}}\left( \tau \right) \right\vert ^{2}+\left\vert \frac{d%
\mathcal{J}_{\mathcal{M}_{S}}\left( \tau \right) }{d\tau }\right\vert ^{2}}{%
\left\vert \mathcal{J}_{\mathcal{M}_{S}}\left( 0\right) \right\vert
^{2}+\left\vert \left. \frac{d\mathcal{J}_{\mathcal{M}_{S}}\left( \tau
\right) }{d\tau }\right\vert _{\tau =0}\right\vert ^{2}}\right) \right] ,
\label{lap}
\end{equation}%
characterizes the exponential growth rate of average statistical volumes
(see (\ref{AVE})) in $\mathcal{M}_{s}$. This suggests that $\lambda _{%
\mathcal{M}_{S}}$ may play the same role as the standard Lyapunov exponent 
\cite{cafaro-phd}. Lyapunov exponents are asymptotic quantities since they
are defined in the limit as time approaches infinity.

\subsection{Weyl Projective Curvature Tensor}

The Weyl Projective curvature tensor \cite{goldberg} (or the anisotropy
tensor) $\mathcal{W}_{abcd}$ is defined as%
\begin{equation}
\mathcal{W}_{abcd}=\mathcal{R}_{abcd}-\frac{\mathcal{R}_{\mathcal{M}_{s}}}{%
N\left( N-1\right) }\left( g_{bd}g_{ac}-g_{bc}g_{ad}\right) ,  \label{Weyl}
\end{equation}%
where $N=nl$ is the dimension of the curved manifold. In (\ref{Weyl}), the
quantity $\mathcal{R}_{abcd}$ is the Riemann curvature tensor defined in the
usual manner by%
\begin{equation}
\mathcal{R}_{\text{ }bcd}^{a}=\partial _{c}\Gamma _{bd}^{a}-\partial
_{d}\Gamma _{bc}^{a}+\Gamma _{fc}^{a}\Gamma _{bd}^{f}-\Gamma _{fd}^{a}\Gamma
_{bc}^{f}.
\end{equation}%
Considerations regarding the negativity of the Ricci curvature as a strong
criterion of dynamical instability and the necessity of compactness of $%
\mathcal{M}_{s}$\textit{\ }in \textquotedblleft true\textquotedblright\
chaotic dynamical systems requires additional investigation.

\subsection{Information Geometric Complexity}

Once the distances among probability distributions have been assigned, a
natural next step is to obtain measures for extended regions in the space of
distributions. Consider an $N$-dimensional volume of the statistical
manifold $\mathcal{M}_{s}$ of distributions $P\left( X|\Theta \right) $
labelled by parameters $\vartheta ^{a}$ with $a=1,\ldots ,N$. The parameters 
$\vartheta ^{a}$ are coordinates for the point $P$ and in these coordinates
it may not be obvious how to write an expression for a volume element $dV_{%
\mathcal{M}_{s}}$. However, within a sufficiently small region (volume
element) any curved space looks flat. Curved spaces are \textquotedblleft
locally flat\textquotedblright . The idea then is rather simple: within that
very small region, we should use Cartesian coordinates and the metric takes
a very simple form, namely the identity matrix $\delta _{ab}$. In locally
Cartesian coordinates $\chi ^{a}$ the volume element is simply given by the
product%
\begin{equation}
dV_{\mathcal{M}_{s}}=d\chi ^{1}d\chi ^{2}\cdots d\chi ^{N},
\end{equation}%
which, in terms of the old coordinates is%
\begin{equation}
dV_{\mathcal{M}_{s}}=\left\vert \frac{\partial \chi }{\partial \Theta }%
\right\vert d\vartheta ^{1}d\vartheta ^{2}\cdots d\vartheta ^{N}=\left\vert 
\frac{\partial \chi }{\partial \Theta }\right\vert d^{N}\Theta .
\end{equation}%
The problem at hand is the calculation of the Jacobian $\left\vert \frac{%
\partial \chi }{\partial \Theta }\right\vert $ of the transformation that
takes the metric $g_{ab}$ into its Euclidean form $\delta _{ab}$.

Let the new coordinates be defined by $\chi ^{a^{\prime }}=\Xi ^{a^{\prime
}}\left( \vartheta ^{1},\ldots ,\vartheta ^{N}\right) $. A small change $%
d\Theta $ corresponds to a small change $d\chi $,%
\begin{equation}
d\chi ^{a^{\prime }}=X_{a}^{a^{\prime }}d\vartheta ^{a}\text{ where }%
X_{a}^{a^{\prime }}\overset{\text{def}}{=}\frac{\partial \chi ^{a^{\prime }}%
}{\partial \vartheta ^{a}},
\end{equation}%
and the Jacobian is given by the determinant of the matrix $X_{a}^{a^{\prime
}}$,%
\begin{equation}
\left\vert \frac{\partial \chi }{\partial \Theta }\right\vert =\left\vert
\det \left( X_{a}^{a^{\prime }}\right) \right\vert .
\end{equation}%
The distance between two neighboring points is the same whether we compute
it in terms of the old or the new coordinates,%
\begin{equation}
dl^{2}=g_{ab}d\vartheta ^{a}d\vartheta ^{b}=\delta _{a^{\prime }b^{\prime
}}d\chi ^{a^{\prime }}d\chi ^{b^{\prime }}.
\end{equation}%
Therefore, the relation between the old and the new metric is%
\begin{equation}
g_{ab}=\delta _{a^{\prime }b^{\prime }}X_{a}^{a^{\prime }}X_{b}^{b^{\prime
}}.  \label{metric relation}
\end{equation}%
Taking the determinant of (\ref{metric relation}), we obtain%
\begin{equation}
g\overset{\text{def}}{=}\det \left( g_{ab}\right) =\left[ \det \left(
X_{a}^{a^{\prime }}\right) \right] ^{2}
\end{equation}%
and therefore%
\begin{equation}
\left\vert \det \left( X_{a}^{a^{\prime }}\right) \right\vert =\sqrt{g}.
\end{equation}%
Finally, we have succeeded in expressing the volume element totally in terms
of the coordinates $\Theta $ and the known metric $g_{ab}\left( \Theta
\right) $,%
\begin{equation}
dV_{\mathcal{M}_{s}}=\sqrt{g}d^{N}\Theta .
\end{equation}%
The volume of any extended region on the manifold is given by%
\begin{equation}
V_{\mathcal{M}_{s}}=\int dV_{\mathcal{M}_{s}}=\int \sqrt{g}d^{N}\Theta .
\end{equation}%
Observe that $\sqrt{g}d^{N}\Theta $ is a scalar quantity and therefore is
invariant under general coordinate transformations $\Theta \rightarrow
\Theta ^{\prime }$, preserving orientation. The square root of the metric
tensor transforms according to%
\begin{equation}
\sqrt{g\left( \Theta \right) }\overset{\Theta \rightarrow \Theta ^{\prime }}{%
\rightarrow }\left\vert \frac{\partial \Theta ^{\prime }}{\partial \Theta }%
\right\vert \sqrt{g\left( \Theta ^{\prime }\right) },  \label{pre1}
\end{equation}%
and the flat infinitesimal volume element $d^{N}\Theta $ transforms as 
\begin{equation}
d^{N}\Theta \overset{\Theta \rightarrow \Theta ^{\prime }}{\rightarrow }%
\left\vert \frac{\partial \Theta }{\partial \Theta ^{\prime }}\right\vert
d^{N}\Theta ^{\prime }.  \label{pre2}
\end{equation}%
Thus, from (\ref{pre1}) and (\ref{pre2}) we obtain%
\begin{equation}
\sqrt{g\left( \Theta \right) }d^{N}\Theta \overset{\Theta \rightarrow \Theta
^{\prime }}{\rightarrow }\sqrt{g\left( \Theta ^{\prime }\right) }d^{N}\Theta
^{\prime }.  \label{pre3}
\end{equation}%
Equation (\ref{pre3}) implies that the infinitesimal statistical volume
element is invariant under general coordinate transformations that preserve
orientation, that is with positive Jacobian.

The volume of an extended region $\Delta V_{\mathcal{M}_{s}}\left( \tau ;%
\vec{\varsigma}\right) $ of $\mathcal{M}_{s}$ is defined by%
\begin{equation}
\Delta V_{\mathcal{M}_{s}}\left( \tau ;\vec{\varsigma}\right) \overset{\text{%
def}}{=}V_{\mathcal{M}_{s}}\left( \tau ;\vec{\varsigma}\right) -V_{\mathcal{M%
}_{s}}\left( 0;\vec{\varsigma}\right) \overset{\text{def.}}{=}\overset{%
\Theta \left( \tau ;\vec{\varsigma}\right) }{\underset{\Theta \left( 0;\vec{%
\varsigma}\right) }{\int }}\sqrt{g}d^{N}\Theta ,  \label{VER}
\end{equation}%
where\textbf{\ }$\Theta \left( \tau ;\vec{\varsigma}\right) $ are solutions
of the geodesic equations\textbf{\ }(\ref{geodesic equations}) and $\vec{%
\varsigma}=\left( \varsigma ^{1},\ldots ,\varsigma ^{N}\right) $ is the
quantity parameterizing the family of geodesics $\left\{ \vartheta _{%
\mathcal{M}_{s}}^{a}\left( \tau ;\vec{\varsigma}\right) \right\} _{\vec{%
\varsigma}\in \mathbb{R}^{+N}}^{a=1,\ldots ,N}$. The quantity that encodes
relevant information about the stability of neighboring volume elements is
the average volume (information geometric complexity) $\mathcal{C}_{\mathcal{%
M}_{s}}\left( \tau ;\vec{\varsigma}\right) $ defined as \cite{ca-1} 
\begin{equation}
\mathcal{C}_{\mathcal{M}_{s}}\left( \tau ;\vec{\varsigma}\right) \overset{%
\text{def}}{=}\left\langle \Delta V_{\mathcal{M}_{s}}\left( \tau ;\vec{%
\varsigma}\right) \right\rangle _{\tau }\overset{\text{def}}{=}\frac{1}{\tau 
}\dint\limits_{0}^{\tau }\Delta V_{\mathcal{M}_{s}}\left( \tau ^{\prime };%
\vec{\varsigma}\right) d\tau ^{\prime }.  \label{AVE}
\end{equation}%
We will call $\mathcal{C}_{\mathcal{M}_{s}}\left( \tau ;\vec{\varsigma}%
\right) $ the information geometric complexity of the maximally probable
trajectories $\Theta \left( \tau ;\vec{\varsigma}\right) $. The IGC $%
\mathcal{C}_{\mathcal{M}_{s}}\left( \tau ;\vec{\varsigma}\right) $\textbf{\ }%
in (\ref{AVE}) represents the temporal average of the\textbf{\ }$N$\textbf{-}%
fold integral over maximum probability trajectories (geodesics) and serves
as a measure of the number of the accessible macrostates in the
configuration (statistical) manifold $\mathcal{M}_{s}$ after a finite
temporal increment\textbf{\ }$\tau $. In other words, $\mathcal{C}_{\mathcal{%
M}_{s}}\left( \tau ;\vec{\varsigma}\right) $ can be interpreted as the
temporal evolution of the system's uncertainty volume $\mathcal{C}_{\mathcal{%
M}_{s}}\left( 0;\vec{\varsigma}\right) $. For instance, $\mathcal{C}_{%
\mathcal{M}_{s}}\left( 0;\vec{\varsigma}\right) $ may be a spherical volume
of initial points whose center is a given point on an attractor and whose
surface consists of configuration points from nearby trajectories. An
attractor is a subset of the manifold $\mathcal{M}_{s}$\textbf{\ }toward
which almost all sufficiently close trajectories converge asymptotically,
covering it densely as time goes on. Strange attractors are called chaotic
attractors. Chaotic attractors have at least one finite positive Lyapunov
exponent \cite{tel}. As the center of $\mathcal{C}_{\mathcal{M}_{s}}\left( 0;%
\vec{\varsigma}\right) $ and its surface points evolve in time, the
spherical volume becomes an ellipsoid with principal axes in the directions
of contraction and expansion. The average rates of expansion and contraction
along the principal axes are the Lyapunov exponents \cite{wolfA}.

\subsection{Information Geometric Entropy}

Finally, the asymptotic regime of diffusive evolution describing the
possible exponential increase of average volume elements on $\mathcal{M}_{s}$
provides another useful indicator of dynamical chaoticity. The exponential
instability characteristic of chaos forces the system to rapidly explore
large areas (volumes) of $\mathcal{M}_{s}$. We remark that the exponential
instability does not necessarily imply that trajectories explore large areas
(volumes) of $\mathcal{M}_{s}$. The folding mechanism is responsible for a
dense exploration of areas (volumes) that may be small in principle. It is
interesting to note that this asymptotic behavior appears also in the
conventional description of quantum chaos where the von Neumann entropy
increases linearly at a rate determined by the Lyapunov exponents. The
linear increase of entropy as a quantum chaos criterion was introduced by
Zurek and Paz \cite{zurek1}. In our information-geometric approach a
relevant quantity that may be useful to study the degree of instability
characterizing ED models is the information geometrodynamical entropy (IGE)
defined as \cite{ca-1, ca-2, ca-3, ca-4, ca-5, ca-6, ca-7}%
\begin{equation}
\mathcal{S}_{\mathcal{M}_{s}}\left( \tau ;\vec{\varsigma}\right) =\mathcal{S}%
_{\mathcal{M}_{s}}\left[ \Theta \left( \tau ;\vec{\varsigma}\right) ,\Theta
\left( 0;\vec{\varsigma}\right) \right] =\lim_{\tau \rightarrow \infty }\ln
\left\{ \frac{1}{\tau }\dint\limits_{0}^{\tau }d\tau ^{\prime }\left[ 
\overset{\Theta \left( \tau ;\vec{\varsigma}\right) }{\underset{\Theta
\left( 0;\vec{\varsigma}\right) }{\int }}\sqrt{g}d^{N}\Theta \right]
\right\} ,  \label{IGE}
\end{equation}%
where $g=\left\vert \det \left( g_{ab}\right) \right\vert $. More
synthetically,%
\begin{equation}
\mathcal{S}_{\mathcal{M}_{s}}\left( \tau ;\vec{\varsigma}\right) \overset{%
\text{def}}{=}\lim_{\tau \rightarrow \infty }\ln \mathcal{C}_{\mathcal{M}%
_{s}}\left( \tau ;\vec{\varsigma}\right)  \label{IGE1}
\end{equation}%
with $\mathcal{C}_{\mathcal{M}_{s}}\left( \tau ;\vec{\varsigma}\right) $
being the information geometric complexity defined in (\ref{AVE}). The IGE\
is intended to capture the temporal complexity (chaoticity) of ED\ models on
curved statistical manifolds\textbf{\ }$\mathcal{M}_{s}$ by considering the
asymptotic temporal behaviors of the average statistical volumes occupied by
the evolving macrovariables labelling points on $\mathcal{M}_{s}$.

\section{Applications}

In this Section we present five applications of the IGAC. First, we study
the chaotic behavior of an ED Gaussian model describing an arbitrary system
of $l$ uncorrelated degrees of freedom and show that the hyperbolicity of
the non-maximally symmetric $2l$-dimensional statistical manifold $\mathcal{M%
}_{s}$ underlying such an ED Gaussian model leads to linear IGE growth and
to exponential divergence of the Jacobi vector field intensity \cite{ca-6}.
Second, we study the asymptotic behavior of the dynamical complexity of the
maximum probability trajectories on Gaussian statistical manifolds in
presence of correlation-like terms between the macrovariables labeling the
macrostates of the system under investigation. In presence of
correlation-like terms, we observe a power law decay of the information
geometric complexity at a rate determined by the correlation coefficient 
\cite{CM}. We also present an information-geometric analogue of the
Zurek-Paz quantum chaos criterion of linear entropy growth \cite{zurek1}.
This analogy is presented by studying the information geometrodynamics of an
ensemble of macroscopic, random frequency, inverted harmonic oscillators 
\cite{ca-2, ca-5}. Next, we apply the IGAC to study the entropic dynamics on
curved statistical manifolds induced by classical probability distributions
in common use in the study of regular and chaotic quantum energy level
statistics. In doing so, we suggest an information-geometric
characterization of regular and chaotic quantum energy level statistics \cite%
{ca-3, ca-7}. Finally, we apply the IGAC to characterize the quantum
entanglement produced by a head-on collision between two Gaussian wave
packets interacting via a scattering process \cite{PLA, Wang}.

\subsection{Gaussian Statistical Model in Absence of Correlations}

As a first example, we apply the IGAC to study the dynamics of a system with 
$l$ degrees of freedom, each one described by two pieces of relevant
information, its mean expected value and its variance (Gaussian statistical
macrostates). The line element $ds^{2}=g_{ab}\left( \Theta \right)
d\vartheta ^{a}d\vartheta ^{b}$ ($a,b=1,\ldots ,2l$) on $\mathcal{M}_{s}$ is
defined by \cite{ca-6}%
\begin{equation}
ds^{2}=\dsum\limits_{k=1}^{2l}\left( \frac{1}{\sigma _{k}^{2}}d\mu _{k}^{2}+%
\frac{2}{\sigma _{k}^{2}}d\sigma _{k}^{2}\right) .
\end{equation}%
This leads to consider an ED model on a non-maximally symmetric $2l$%
-dimensional statistical manifold $\mathcal{M}_{s}$. Manifold $\mathcal{M}%
_{s}$ possesses a constant negative Ricci curvature that is proportional to
the number of degrees of freedom of the system, $R_{\mathcal{M}_{s}}=-l$.
The system explores statistical volume elements on $\mathcal{M}_{s}$ at an
exponential rate, while the information geometrodynamical entropy $\mathcal{S%
}_{\mathcal{M}_{s}}$\ increases linearly in time (statistical evolution
parameter) and is proportional to the number of degrees of freedom of the
system, $\mathcal{S}_{\mathcal{M}_{s}}$ $\overset{\tau \rightarrow \infty }{%
\sim }l\lambda \tau $. The parameter $\lambda $ characterizes the family of
probability distributions on $\mathcal{M}_{s}$. For the case being
considered here\textbf{,} $\lambda $\textbf{\ }does indeed play the role of
the standard Lyapunov exponent \cite{cafaro-phd}.

Recall that the finite Lyapunov exponent in the direction $v\in 
\mathbb{R}
^{2l}$ of a trajectory $\Theta \left( \tau ,\Theta _{0}\right) $ satisfying
the differential equation $\dot{\Theta}=\mathcal{A}\left( \tau \right)
\Theta $ with $\Theta \in 
\mathbb{R}
^{2l}$ and initial condition $\Theta \left( 0,\Theta _{0}\right) =\Theta _{0}
$ is defined as \cite{lyapunov}%
\begin{equation}
\lambda \left( v\right) \overset{\text{def}}{=}\lim_{\tau \rightarrow \infty
}\ln \left[ \frac{\sqrt{\left\langle \mathcal{X}v,\mathcal{X}v\right\rangle }%
}{\sqrt{\left\langle v,v\right\rangle }}\right] .  \label{111}
\end{equation}%
The brackets $\left\langle \cdot ,\cdot \right\rangle $ in (\ref{111})
denote the standard scalar product in $%
\mathbb{R}
^{2l}$ and $\mathcal{X}=\mathcal{X}\left( \tau ;\Theta \left( \tau ,\Theta
_{0}\right) \right) $ is the asymptotically regular fundamental matrix of
the differential equation $\dot{\Theta}-\mathcal{A}\left( \tau \right)
\Theta =0$ \cite{verhulst}. For instance, in the $2l$-dimensional Gaussian
statistical model considered here, the set of differential equations to
consider is%
\begin{equation}
\frac{d\Theta \left( \tau \right) }{d\tau }-\mathcal{A}\left( \tau \right)
\Theta \left( \tau \right) =0,
\end{equation}%
where $\Theta \left( \tau \right) $ is the $2l$-dimensional vector $\Theta
\left( \tau \right) \equiv \left( \mu _{1}\left( \tau \right) ,\ldots ,\mu
_{l}\left( \tau \right) ;\sigma _{1}\left( \tau \right) ,\ldots ,\sigma
_{l}\left( \tau \right) \right) $ whose components are solutions of geodesic
equations describing the evolution of the macrostates of the system \cite%
{ca-6}. In the asymptotic limit, the $2l\times 2l$ matrix $\mathcal{A}\left(
\tau \right) $ can be approximated by a diagonal matrix with constant
coefficients, $\mathcal{A}\left( \tau \right) \overset{\tau \rightarrow
\infty }{\approx }$ $\mathrm{diag}\left( 0,\lambda _{1},0,\lambda
_{2},\ldots ,0,\lambda _{l}\right) $. A straightforward calculation leads to
an asymptotically regular $2l\times 2l$ fundamental matrix 
\begin{equation}
\mathcal{X}\left( \tau \right) \overset{\tau \rightarrow \infty }{\approx }%
\mathrm{diag}\left( c_{1}\tau ,c_{2}\exp \left( \lambda _{1}\tau \right)
,c_{3}\tau ,c_{4}\exp \left( \lambda _{2}\tau \right) ,\ldots ,c_{l-1}\tau
,c_{l}\exp \left( \lambda _{l}\tau \right) \right) ,
\end{equation}%
with $c_{i}\in 
\mathbb{R}
$, $\forall i=1,\ldots ,l$. Therefore, equation (\ref{111}) leads to the
following interesting result,%
\begin{equation}
\lambda _{\max }\left( v\right) =\underset{%
\mathbb{R}
^{+}}{\max }\left\{ \lambda _{1},\ldots ,\lambda _{l}\right\} ,\text{ }%
\forall v\in 
\mathbb{R}
^{2l}.
\end{equation}%
Thus, the quantities $\lambda _{k}$ with $k=1,\ldots ,l$ are indeed Lyapunov
exponents. In this case, for the sake of simplicity, we have assumed $%
\lambda _{i}=\lambda _{j}$ $\forall i,$ $j=1,\ldots ,l$. The asymptotic
linear information-geometrodynamical entropy growth of $\mathcal{S}_{%
\mathcal{M}_{s}}\left( \tau \right) $ may be considered the
information-geometric analogue of the von Neumann entropy growth introduced
by Zurek-Paz, a \textit{quantum} feature of chaos.

The geodesics on $\mathcal{M}_{s}$ are hyperbolic trajectories. Using the
Jacobi-Levi-Civita (JLC) equation for geodesic spread, we show that the
Jacobi vector field intensity $\mathcal{J}_{\mathcal{M}_{s}}$ diverges
exponentially and is proportional to the number of degrees of freedom of the
system, $\mathcal{J}_{\mathcal{M}_{s}}$ $\overset{\tau \rightarrow \infty }{%
\sim }l\exp \left( \lambda \tau \right) $. The exponential divergence of the
Jacobi vector field intensity $\mathcal{J}_{\mathcal{M}_{s}}$ is a \textit{%
classical} feature of chaos. Therefore, we conclude that \cite{ca-6} 
\begin{equation}
\mathcal{R}_{\mathcal{M}_{s}}=-l,\text{ }\mathcal{J}_{\mathcal{M}_{s}}%
\overset{\tau \rightarrow \infty }{\sim }l\exp \left( \lambda \tau \right) ,%
\text{ }\mathcal{S}_{\mathcal{M}_{s}}\overset{\tau \rightarrow \infty }{\sim 
}l\lambda \tau .  \label{link}
\end{equation}%
By virtue of (\ref{link}) we observe that $\mathcal{R}_{\mathcal{M}_{s}}$, $%
\mathcal{S}_{\mathcal{M}_{s}}$ and $\mathcal{J}_{\mathcal{M}_{s}}$ are
proportional to the number of Gaussian-distributed microstates of the
system. This proportionality, even though proven in a very special case,
suggest there may be a substantial link among these information-geometric
indicators of chaoticity.

\subsection{Gaussian Statistical Model in presence of Macro-correlations}

As a second example, we apply the IGAC to study the information constrained
dynamics of a system with $l$ degrees of freedom, each one described by two
\textquotedblleft correlated\textquotedblright\ pieces of relevant
information, its mean expected value and its variance (Gaussian statistical
macrostates). The line element $ds^{2}=g_{ab}\left( \Theta \right)
d\vartheta ^{a}d\vartheta ^{b}$ ($a,b=1,\ldots ,2l$) on the $2l$-dimensional
Gaussian statistical manifold $\mathcal{M}_{s}$ in the presence of non
trivial off-diagonal terms is given by%
\begin{equation}
ds_{\mathcal{M}_{s}}^{2}=\sum_{j=1}^{2l}\left( \frac{1}{\sigma _{j}^{2}}d\mu
_{j}^{2}+\frac{2r_{j}}{\sigma _{j}^{2}}d\mu _{j}d\sigma _{j}+\frac{2}{\sigma
_{j}^{2}}d\sigma _{j}^{2}\right) .  \label{SM}
\end{equation}%
We consider positive coefficients $r_{j}\in \left( 0,1\right) $, $\forall
j=1,\ldots ,l$. \ From (\ref{SM}), it can be shown that the Ricci scalar
curvature $\mathcal{R}_{\mathcal{M}_{s}}\left( r_{1},\ldots ,r_{l}\right) $
of such a $2l$-dimensional manifold is given by%
\begin{equation*}
\mathcal{R}_{\mathcal{M}_{s}}\left( r_{1},\ldots ,r_{l}\right)
=-2^{3}\sum_{k=1}^{2l}\left( 2-r_{k}^{2}\right) ^{-3}.
\end{equation*}%
Notice that in the limit of vanishing correlation strengths $\left\{
r_{k}\right\} $, $\mathcal{R}_{\mathcal{M}_{s}}=-l$ as shown in \cite{ca-6}.
Applying the IGAC formalism, we are able to compute the asymptotic temporal
behavior of the dynamical complexity of geodesic trajectories for the
correlated $2l$-dimensional Gaussian statistical model. The technical
details that will be omitted in what follows may be found in \cite{CM}. It
turns out that \cite{CM}%
\begin{equation}
\mathcal{S}_{\mathcal{M}_{s}}\left( \tau ;\left\{ \lambda _{k}\right\}
,\left\{ r_{k}\right\} \right) \overset{\tau \rightarrow \infty }{\sim }%
\sum_{k=1}^{2l}\ln \left[ \Lambda _{1}\left( r_{k}\right) +\frac{\Lambda
_{2}\left( r_{k},\lambda _{k}\right) }{\tau }\right] ,
\end{equation}%
where%
\begin{equation}
\Lambda _{1}\left( r_{k}\right) \overset{\text{def}}{=}\frac{2r_{k}\sqrt{%
2-r_{k}^{2}}}{1+\sqrt{1+4r_{k}^{2}}},\text{ }\Lambda _{2}\left(
r_{k},\lambda _{k}\right) \overset{\text{def}}{=}\frac{\sqrt{\left(
1+4r_{k}^{2}\right) \left( 2-r_{k}^{2}\right) }}{r_{k}}\frac{\ln \Sigma
\left( r_{k},\lambda _{k},\alpha _{\pm }\right) }{\lambda _{k}},\text{ }%
\alpha _{\pm }\left( r_{k}\right) \overset{\text{def}}{=}\frac{3\pm \sqrt{%
1+4r_{k}^{2}}}{2}.\text{ }
\end{equation}%
The quantity $\Sigma \left( r_{k},\lambda _{k},\alpha _{\pm }\right) $ is a
strictly positive function of its arguments. For $r_{k}=r_{s}$ $\forall k$
and $s=1,\ldots ,l$, the information geometric entropy $\mathcal{S}_{%
\mathcal{M}_{s}}\left( \tau ;l,\lambda ,r\right) $ becomes%
\begin{equation}
\mathcal{S}_{\mathcal{M}_{s}}\left( \tau ;l,\lambda ,r\right) \overset{\tau
\rightarrow \infty }{\sim }\ln \left[ \Lambda _{1}\left( r\right) +\frac{%
\Lambda _{2}\left( r,\lambda \right) }{\tau }\right] ^{l}.
\end{equation}%
In this case, it is clear that the IGE presents a power law decay, where the
power is related to the cardinality $l$ of the microscopic degrees of
freedom characterized by correlated pieces of macroscopic information.
Furthermore, the IGE reaches a saturation value quantified by the set $%
\left\{ r_{k}\right\} $ (the correlation strengths). It appears that
macro-correlations lead to the emergence of an asymptotic information
geometric compression of the explored statistical macrostates on the
configuration manifold of the model in its evolution between initial and
final macrostates $\Theta ^{\left( \text{initial}\right) }\left( 0\right) $
and $\Theta ^{\left( \text{final}\right) }\left( \tau \right) $,
respectively.

\subsection{Ensemble of Random Frequency Macroscopic Inverted Harmonic
Oscillators}

For our\textbf{\ third} example, we employ ED in conjunction with
\textquotedblleft Newtonian Entropic Dynamics\textquotedblright\ (NED) \cite%
{cafaro4}. In NED, we explore the possibility of using well established
principles of inference to derive Newtonian dynamics from relevant prior
information codified into an appropriate statistical manifold. The basic
assumption is that there is an irreducible uncertainty in the location of
particles so that the position of a particle is defined by a probability
distribution. The corresponding configuration space is a statistical
manifold $\mathcal{M}_{s}$ the geometry of which is defined by the
Fisher-Rao information metric. The trajectory follows from a principle of
inference, namely the method of Maximum relative Entropy. There is no need
for additional \textquotedblleft physical\textquotedblright\ postulates such
as an action principle or equation of motion, nor for the concept of mass,
momentum or phase space, not even the notion of time. The resulting
\textquotedblleft entropic\textquotedblright\ dynamics reproduces Newton's
mechanics for any number of particles interacting among themselves and with
external fields. Both the mass of the particles and their interactions are
explained as a consequence of the underlying statistical manifold.

In our special application, we consider a manifold with a line element $%
ds^{2}=g_{ab}\left( \Theta \right) d\vartheta ^{a}d\vartheta ^{b}$ ($%
a,b=1,\ldots ,l$) given by \cite{ca-2, ca-5}%
\begin{equation}
ds^{2}=\left[ 1-\Phi \left( \Theta \right) \right] \delta _{ab}\left( \Theta
\right) d\vartheta ^{a}d\vartheta ^{b},\text{ }\Phi \left( \Theta \right) =%
\overset{l}{\underset{k=1}{\sum }}u_{k}\left( \vartheta _{k}\right) ,\text{ }
\end{equation}%
where%
\begin{equation}
u_{k}\left( \theta _{k}\right) =-\frac{1}{2}\omega _{k}^{2}\vartheta
_{k}^{2},\text{ }\vartheta _{k}=\vartheta _{k}\left( s\right) .
\end{equation}%
The geodesic equations for the macrovariables $\vartheta _{k}\left( s\right) 
$ are strongly \textit{nonlinear }and their integration is not trivial.
However, upon a suitable change of the affine parameter $s$ used in the
geodesic equations, we may simplify the differential equations for the
macroscopic variables parametrizing points on the manifold $\mathcal{M}_{s}%
\overset{\text{def}}{=}\mathcal{M}_{\text{IHO}}^{\left( l\right) }$ with
metric tensor $g_{ab}$. Recalling that the notion of chaos is
observer-dependent and upon changing the affine parameter from $s$ to $\tau $
in such a way that $ds^{2}=2\left( 1-\Phi \right) ^{2}d\tau ^{2}$, we obtain
new geodesic equations describing a set of macroscopic inverted harmonic
oscillators (IHOs). In this example, the IGE $S_{\mathcal{M}_{\text{IHO}%
}^{\left( l\right) }}\left( \tau ;\omega _{1},\ldots ,\omega _{l}\right) $
reads%
\begin{equation}
S_{\mathcal{M}_{\text{IHO}}^{\left( l\right) }}\left( \tau ;\omega
_{1},\ldots ,\omega _{l}\right) \overset{\text{def}}{=}\lim_{\tau
\rightarrow \infty }\ln \mathcal{C}_{\mathcal{M}_{\text{IHO}}^{\left(
l\right) }}\left( \tau ;\omega _{1},\ldots ,\omega _{l}\right) ,
\label{gen-ent}
\end{equation}%
where%
\begin{equation}
\mathcal{C}_{\mathcal{M}_{\text{IHO}}^{\left( l\right) }}\left( \tau ;\omega
_{1},\ldots ,\omega _{l}\right) =\frac{1}{\tau }\dint\limits_{0}^{\tau
}\Delta V_{\mathcal{M}_{\text{IHO}}^{\left( l\right) }}\left( \tau ^{\prime
};\omega _{1},\ldots ,\omega _{l}\right) d\tau ^{\prime },  \label{inter2}
\end{equation}%
and%
\begin{equation}
\Delta V_{\mathcal{M}_{\text{IHO}}^{\left( l\right) }}\left( \tau ^{\prime
};\omega _{1},\ldots ,\omega _{l}\right) =\underset{\left\{ \vec{\vartheta}%
^{\prime }\right\} }{\int }d^{l}\vec{\vartheta}^{\prime }\left( 1+\frac{1}{2}%
\underset{j=1}{\overset{l}{\sum }}\omega _{j}^{2}\vartheta _{j}^{\prime
2}\right) ^{\frac{l}{2}}.  \label{inter3}
\end{equation}%
Substituting (\ref{inter2}) and (\ref{inter3}) in (\ref{gen-ent}), we obtain
the general expression for $S_{\mathcal{M}_{\text{IHO}}^{\left( l\right)
}}\left( \tau ;\omega _{1},\ldots ,\omega _{l}\right) $, 
\begin{equation}
S_{\mathcal{M}_{\text{IHO}}^{\left( l\right) }}\left( \tau ;\omega
_{1},\ldots ,\omega _{l}\right) \overset{\text{def}}{=}\lim_{\tau
\rightarrow \infty }\ln \left\{ \frac{1}{\tau }\int_{0}^{\tau }\left[ 
\underset{\left\{ \vec{\vartheta}^{\prime }\right\} }{\int }d^{l}\vec{%
\vartheta}^{\prime }\left( 1+\frac{1}{2}\underset{j=1}{\overset{l}{\sum }}%
\omega _{j}^{2}\vartheta _{j}^{\prime 2}\right) ^{\frac{l}{2}}\right] d\tau
^{\prime }\right\} .  \label{inter4}
\end{equation}%
To evaluate (\ref{inter4}) we observe that $\Delta V_{\mathcal{M}_{\text{IHO}%
}^{\left( l\right) }}$ in (\ref{inter3}) can be written as%
\begin{equation}
\Delta V_{\mathcal{M}_{\text{IHO}}^{\left( l\right) }}\left( \tau ^{\prime
};\omega _{1},\ldots ,\omega _{l}\right) \overset{\text{ }}{\approx }\frac{1%
}{l}\frac{1}{2^{\frac{l}{2}}}\left( \overset{l}{\underset{i=1}{\Pi }}%
\vartheta _{i}^{\prime }\right) \left[ \underset{j=1}{\overset{l}{\sum }}%
\omega _{j}^{2}\vartheta _{j}^{\prime 2}\right] ^{\frac{l}{2}}.
\end{equation}%
Since the $l$-Newtonian equations of motions for each IHO are given by%
\begin{equation}
\frac{d^{2}\vartheta _{j}}{d\tau ^{2}}-\omega _{j}^{2}\vartheta _{j}=0,\text{
}\forall j=1,\ldots ,l,
\end{equation}%
the asymptotic behavior of such macrovariables on manifold $\mathcal{M}_{%
\text{IHO}}^{\left( l\right) }$ is given by%
\begin{equation}
\vartheta _{j}\left( \tau \right) \overset{\tau \rightarrow \infty }{\approx 
}\Xi _{j}e^{\omega _{j}\tau },\text{ }\Xi _{j}\in 
\mathbb{R}
,\text{ }\forall j=1,\ldots ,l.
\end{equation}%
We therefore obtain%
\begin{equation}
\Delta V_{\mathcal{M}_{\text{IHO}}^{\left( l\right) }}\left( \tau ;\omega
_{1},\ldots ,\omega _{l}\right) \overset{\tau \rightarrow \infty \text{ }}{%
\approx }\frac{1}{l}\frac{1}{2^{\frac{l}{2}}}\left( \underset{i=1}{\overset{l%
}{\Pi }}\Xi _{i}\right) \cdot \exp \left( \overset{l}{\underset{i=1}{\sum }}%
\omega _{i}\tau \right) \left[ \underset{j=1}{\overset{l}{\sum }}\Xi
_{j}^{2}e^{2\omega _{j}\tau }\omega _{j}^{2}\right] ^{\frac{l}{2}}.
\label{inter5}
\end{equation}%
Upon averaging (\ref{inter5}), we find%
\begin{equation}
\mathcal{C}_{\mathcal{M}_{\text{IHO}}^{\left( l\right) }}\left( \tau ;\omega
_{1},\ldots ,\omega _{l}\right) \overset{\tau \rightarrow \infty \text{ }}{%
\approx }\frac{1}{\tau }\dint\limits_{0}^{\tau }\left\{ \frac{1}{l}\frac{1}{%
2^{\frac{l}{2}}}\left( \underset{i=1}{\overset{l}{\Pi }}\Xi _{i}\right)
\cdot \exp \left( \Omega \tau ^{\prime }\right) \left[ \underset{j=1}{%
\overset{l}{\sum }}\Xi _{j}^{2}e^{2\omega _{j}\tau ^{\prime }}\omega _{j}^{2}%
\right] ^{\frac{l}{2}}\right\} d\tau ^{\prime },
\end{equation}%
where $\Omega =\overset{l}{\underset{i=1}{\sum }}\omega _{i}$. As a working
hypothesis, we assume $\Xi _{i}=\Xi _{j}\equiv \Xi $ $\forall i,$ $%
j=1,\ldots ,l$. Furthermore, assume that $n\rightarrow \infty $ so that the
spectrum of frequencies becomes continuum and as an additional working
hypothesis, assume this spectrum is linearly distributed (Ohmic frequency
spectrum),%
\begin{equation}
\rho _{\text{Ohmic}}\left( \omega \right) =\frac{2}{\Omega _{\text{cut-off}%
}^{2}}\omega \text{ with}\underset{0}{\overset{\Omega _{\text{cut-off}}}{%
\int }}\rho _{\text{Ohmic}}\left( \omega \right) d\omega =1,\text{ }\Omega _{%
\text{cut-off}}=\xi \Omega ,\text{ }\xi \in 
\mathbb{R}
.
\end{equation}%
Thus, we obtain 
\begin{equation}
\mathcal{C}_{\mathcal{M}_{\text{IHO}}^{\left( l\right) }}\left( \tau ;\omega
_{1},\ldots ,\omega _{l}\right) \overset{\tau \rightarrow \infty \text{ }}{%
\approx }\frac{1}{l}\frac{1}{2^{\frac{l}{2}}}\Xi ^{2l}\left( \frac{\xi
^{2}\Omega ^{2}}{2}\right) ^{\frac{l}{2}}\frac{\exp \left( \frac{l}{2}\xi
\Omega \tau \right) }{\tau }.  \label{inter6}
\end{equation}%
Finally, substituting (\ref{inter6}) into (\ref{gen-ent}) yields \cite{ca-3,
ca-7} 
\begin{equation}
\mathcal{S}_{\mathcal{M}_{\text{IHO}}^{\left( l\right) }}\left( \tau ;\omega
_{1},\ldots ,\omega _{l}\right) \overset{\tau \rightarrow \infty }{\propto }%
\Omega \tau ,\text{ }\Omega =\overset{l}{\underset{i=1}{\sum }}\omega _{i}.
\label{Fin}
\end{equation}%
Equation (\ref{Fin}) displays the asymptotic, linear information
geometrodynamical entropy growth of the generalized $l$-set of inverted
harmonic oscillators and extends the result of Zurek-Paz to an arbitrary set
of anisotropic inverted harmonic oscillators \cite{zurek1} in a classical
information-geometric setting. In order to ensure the compactification of
the parameter space of the system (and therefore $\mathcal{M}_{s}$ itself),
it is possible to choose a Gaussian distributed frequency spectrum for the
IHOs. With this choice of frequency spectrum, the folding mechanism required
for true chaos is restored in a statistical (averaging over $\omega $ and $%
\tau $) sense. This example may be considered the information geometric
analogue of the Zurek-Paz model used to investigate the implications of
decoherence for quantum chaos. In their work, Zurek and Paz considered a
chaotic system, a single unstable harmonic oscillator characterized by a
potential $V\left( x\right) =-\frac{\Omega ^{2}x^{2}}{2}$ ($\Omega $ is the
Lyapunov exponent), coupled to an external environment. In the \textit{%
reversible classical limit }\cite{zurek2}, the von Neumann entropy of such a
system increases linearly at a rate determined by the Lyapunov exponent, 
\begin{equation}
\mathcal{S}_{\text{quantum}}^{\left( \text{chaotic}\right) }\left( \tau
\right) \overset{\tau \rightarrow \infty }{\sim }\Omega \tau ,
\end{equation}%
with $\Omega $ playing the role of the Lyapunov exponent.

\subsection{Regular and Chaotic Quantum Spin Chains}

In our fourth example, we apply the IGAC to study the entropic dynamics on
curved statistical manifolds induced by classical probability distributions
commonly used in the study of regular and chaotic quantum energy level
statistics. In doing so, we suggest an information-geometric
characterization of a special class of regular and chaotic quantum energy
level statistics.

Recall that the theory of quantum chaos (quantum mechanics of systems whose
classical dynamics are chaotic) is not primarily related to few-body
physics. Indeed, in real physical systems such as many-electron atoms and
heavy nuclei, the origin of complex behavior is the very strong interaction
among many particles. To deal with such systems, a famous statistical
approach has been developed which is based upon Random Matrix Theory (RMT).
The main idea of this approach is to neglect the detailed description of the
motion and to treat these systems statistically, bearing in mind that the
interaction among particles is so complex and strong that generic properties
are expected to emerge. Once again, this is exactly the philosophy
underlining the ED approach to complex dynamics. It is known that the
asymptotic behavior of computational costs and entanglement entropies of
integrable and chaotic Ising spin chains are very different \cite{prosen}.
Prosen considered the question of time efficiency in implementing an
up-to-date version of the t-DMRG for a family of Ising spin $\frac{1}{2}$
chains in an arbitrarily oriented magnetic field which undergoes a
transition from an integrable (transverse Ising) to nonintegrable chaotic
regime as the magnetic field is varied. An integrable (regular) Ising chain
in a general homogeneous, transverse magnetic field is defined through the
Hamiltonian $\mathcal{H}_{\text{regular}}\left( 0,2\right) $, where%
\begin{equation}
\mathcal{H}\left( h_{x},h_{y}\right) =\underset{j=0}{\overset{n-2}{\sum }}%
\sigma _{j}^{x}\sigma _{j+1}^{x}+\underset{j=0}{\overset{n-1}{\sum }}\left(
h^{x}\sigma _{j}^{x}+h^{y}\sigma _{j}^{y}\right) .  \label{ising-hamiltonian}
\end{equation}%
In this case, the computational cost presents a polynomial growth in time, $%
D_{\varepsilon }^{\left( \text{regular}\right) }\left( t\right) \overset{%
\tau \rightarrow \infty }{\propto }\tau $, while the entanglement entropy is
characterized by logarithmic growth,%
\begin{equation}
\mathcal{S}_{\text{regular}}\left( 0,2\right) =\mathcal{S}_{\text{von Neumann%
}}^{\left( 0,2\right) }\overset{\tau \rightarrow \infty }{\propto }c\ln \tau
+c^{\prime }.  \label{regularentropy}
\end{equation}%
The constant\textbf{\ }$c$\ depends exclusively on the value of the fixed
transverse magnetic field intensity\textbf{\ }$B_{\perp }$, while\textbf{\ }$%
c^{\prime }$\textbf{\ }depends on\textbf{\ }$B_{\perp }$\textbf{\ }and on
the choice of the initial local operators of finite index used to calculate
the operator space entanglement entropy. In contrast, a quantum chaotic
Ising chain in a general homogeneous tilted magnetic field is defined
through the Hamiltonian $\mathcal{H}_{\text{chaotic}}\left( 1,1\right) $,
where $\mathcal{H}$ is given in (\ref{ising-hamiltonian}). In this case, the
computational cost presents an exponential growth in time, $D_{\varepsilon
}^{\left( \text{chaotic}\right) }\left( t\right) \overset{\tau \rightarrow
\infty }{\propto }\exp \left( \mathcal{K}_{q}\tau \right) $, while the
entanglement entropy is characterized by linear growth,%
\begin{equation}
\mathcal{S}_{\text{chaotic}}\left( 1,1\right) =\mathcal{S}_{\text{von Neumann%
}}^{\left( 1,1\right) }\overset{\tau \rightarrow \infty }{\propto }\mathcal{K%
}_{q}\tau .  \label{QDE}
\end{equation}%
The quantity\textbf{\ }$\mathcal{K}_{q}$\textbf{\ }is a constant, is
asymptotically independent of the number of indexes of the initial local
operators used to calculate the operator space entropy, depends only on the
Hamiltonian evolution and not on the details of the initial state observable
or error measures, and can be interpreted as a kind of quantum dynamical
entropy.

It is well known the quantum description of chaos is characterized by a
radical change in the statistics of quantum energy levels \cite{casati}. The
transition to chaos in the classical limit of quantum systems is associated
with a drastic change in the statistics of the nearest-neighbor spacings of
quantum energy levels. In the regular regime, the distribution agrees with
Poisson statistics while in the chaotic regime the Wigner-Dyson distribution
works very well. Uncorrelated energy levels are characteristic of quantum
systems corresponding to a classically regular motion while a level
repulsion (a suppression of small energy level spacing) is typical for
systems which are classically chaotic. A standard quantum example is
provided by the study of energy level statistics of an Hydrogen atom in a
strong magnetic field. It is known that level spacing distribution (LSD) is
a standard indicator of quantum chaos \cite{haake}. It displays
characteristic level repulsion for strongly nonintegrable quantum systems,
whereas for integrable systems there is no repulsion due to the existence of
conservation laws and quantum numbers. In \cite{prosen}, the authors
calculate the LSD of the spectra of $\mathcal{H}_{\text{regular}}\left(
0,2\right) $ and $\mathcal{H}_{\text{chaotic}}\left( 1,1\right) $. They find
that for $\mathcal{H}_{\text{regular}}\left( 0,2\right) $, the nearest
neighbor LSD is described by a Poisson distribution. For $\mathcal{H}_{\text{%
chaotic}}\left( 1,1\right) $, they find the nearest neighbor LSD is
described by a Wigner-Dyson distribution. Therefore, they conclude that $%
\mathcal{H}_{\text{regular}}\left( 0,2\right) $ and $\mathcal{H}_{\text{%
chaotic}}\left( 1,1\right) $ indeed represent generic regular and quantum
chaotic systems, respectively. We encode the relevant information about the
spin-chain in a suitable composite-probability distribution taking account
of the quantum spin chain and the configuration of the external magnetic
field in which they are immersed.

In the ME\ method, the selection of relevant variables is made on the basis
of intuition guided by experiment; it is essentially a matter of trial and
error. The variables should include those that can be controlled or
experimentally observed, but there are cases where others must also be
considered. Our objective here is to choose the relevant microvariables of
the system and select the relevant information concerning each one of them.
In the integrable case, the Hamiltonian $\mathcal{H}_{\text{regular}}\left(
0,2\right) $ describes an antiferromagnetic Ising chain immersed in a
transverse, homogeneous magnetic field $\vec{B}_{\text{transverse}}=B_{\perp
}$ $\hat{B}_{_{\perp }}$ where the level spacing distribution of its
spectrum is given by the Poisson distribution%
\begin{equation}
p_{\text{Poisson}}\left( x_{A}|\mu _{A}\right) =\frac{1}{\mu _{A}}\exp
\left( -\frac{x_{A}}{\mu _{A}}\right) .  \label{poisson-integrable}
\end{equation}%
The microvariable $x_{A}$ represents the spacing of the energy levels while
the macrovariable $\mu _{A}$ is the average spacing. The chain is immersed
in a \textit{transverse} magnetic field which has just one component $%
B_{\perp }$ in the Hamiltonian $\mathcal{H}_{\text{regular}}\left(
0,2\right) $. Observe that the Exponential distribution is identified by
information theory as the maximum entropy distribution if only one piece of
information (the expectation value) is known\textbf{.} Thus, we translate
this piece of information in the IGAC formalism by coupling the probability (%
\ref{poisson-integrable}) to an exponential bath $p_{B}^{\left( \text{%
exponential}\right) }\left( x_{B}|\mu _{B}\right) $ given by%
\begin{equation}
p_{B}^{\left( \text{exponential}\right) }\left( x_{B}|\mu _{B}\right) =\frac{%
1}{\mu _{B}}\exp \left( -\frac{x_{B}}{\mu _{B}}\right) ,
\end{equation}%
where the microvariable $x_{B}$ is the intensity of the magnetic field and
the macrovariable $\mu _{B\text{ }}$is the average intensity. More
correctly, $x_{B}$\ should be the energy arising from the interaction of the 
\textit{transverse} magnetic field with the spin $\frac{1}{2}$\ particle
magnetic moment, $x_{B}=\left\vert -\vec{\mu}\cdot \vec{B}\right\vert
=\left\vert -\mu B\cos \varphi \right\vert $\ where $\varphi $\ is the tilt
angle. For the sake of simplicity, let us set $\mu =1$. Then in the
transverse case $\varphi =0$\ and therefore $x_{B}=B\equiv B_{\perp }$. This
is our best guess and we justify it by noticing that the magnetic field
intensity is indeed a relevant quantity in this experiment (see equation (%
\ref{regularentropy})). Its components are varied during the transition from
integrable to chaotic regimes. In the regular regime, we say the magnetic
field intensity is set to a well-defined value $\left\langle
x_{B}\right\rangle =\mu _{B}$. Finally, the chosen composite probability
distribution $P^{\left( \text{integrable}\right) }\left( x_{A},x_{B}|\mu
_{A},\mu _{B}\right) $ encoding relevant information about the system is
given by%
\begin{equation}
P^{\left( \text{integrable}\right) }\left( x_{A},x_{B}|\mu _{A},\mu
_{B}\right) =\frac{1}{\mu _{A}\mu _{B}}\exp \left[ -\left( \frac{x_{A}}{\mu
_{A}}+\frac{x_{B}}{\mu _{B}}\right) \right] .  \label{p-exp}
\end{equation}%
Again, we point out that our probability (\ref{p-exp}) is our best guess
and, of course, must be consistent with numerical simulations and
experimental data in order to have some merit. We point out that equation (%
\ref{p-exp}) is not fully justified from a theoretical point of view, a
situation that occurs due to the lack of a systematic way to select the
relevant microvariables of the system (and to choose the appropriate
information about such microvariables). Let us denote $\mathcal{M}_{S}^{%
\text{integrable}}$ the two-dimensional curved statistical manifold
underlying our information geometrodynamics. The line element $ds_{\text{%
integrable}}^{2}$ on $\mathcal{M}_{S}^{\text{integrable}}$ is given by%
\begin{equation}
ds_{\text{integrable}}^{2}=ds_{\text{Poisson}}^{2}+ds_{\text{Exponential}%
}^{2}=\frac{1}{\mu _{A}^{2}}d\mu _{A}^{2}+\frac{1}{\mu _{B}^{2}}d\mu
_{B}^{2}.  \label{regular case}
\end{equation}%
Applying the IGAC to the line element in (\ref{regular case}) leads to
conclude polynomial growth in $\mathcal{C}_{\mathcal{M}_{s}}^{\text{%
integrable}}$ and logarithmic IGE growth \cite{ca-3, ca-7}, 
\begin{equation}
\mathcal{C}_{\mathcal{M}_{s}}^{\left( \text{integrable}\right) }\left( \tau
\right) \overset{\tau \rightarrow \infty }{\propto }\exp (c_{\text{IG}%
}^{\prime })\tau ^{c_{\text{IG}}},\text{ }\mathcal{S}_{\mathcal{M}%
_{s}}^{\left( \text{integrable}\right) }\left( \tau \right) \overset{\tau
\rightarrow \infty }{\propto }c_{\text{IG}}\ln \tau +c_{\text{IG}}^{\prime }.
\label{polinomio}
\end{equation}%
The quantity\textbf{\ }$c_{\text{IG}}$\textbf{\ }is a constant proportional
to the number of Exponential probability distributions in the composite
distribution used to calculate the IGE; $c_{\text{IG}}^{\prime }$\ is a
constant that depends on the values assumed by the statistical macrovariables%
\textbf{\ }$\mu _{A}$\textbf{\ }and\textbf{\ }$\mu _{B}$. Equations (\ref%
{polinomio}) may be interpreted as the information-geometric analogue of the
computational complexity $D_{\varepsilon }^{\left( \text{regular}\right)
}\left( \tau \right) $\ and the entanglement entropy $\mathcal{S}_{\text{%
regular}}\left( 0,2\right) $\ defined in standard quantum information
theory, respectively. We cannot state they are the same since we are not
fully justifying, from a theoretical standpoint, our choice of the composite
probability (\ref{p-exp}).

In the chaotic case, the Hamiltonian $\mathcal{H}_{\text{chaotic}}\left(
1,1\right) $ describes an antiferromagnetic Ising chain immersed in a
tilted, homogeneous magnetic field $\vec{B}_{\text{tilted}}=B_{\perp }$ $%
\hat{B}_{\perp }+B_{\parallel }$ $\hat{B}_{\parallel }$, with the level
spacing distribution of its spectrum given by the Poisson distribution $p_{%
\text{Wigner-Dyson}}\left( x_{A}^{\prime }|\mu _{A}^{\prime }\right) $,%
\begin{equation}
p_{\text{Wigner-Dyson}}\left( x_{A}^{\prime }|\mu _{A}^{\prime }\right) =%
\frac{\pi x_{A}^{\prime }}{2\mu _{A}^{\prime 2}}\exp \left( -\frac{\pi
x_{A}^{\prime 2}}{4\mu _{A}^{\prime 2}}\right) ,
\label{wigner-dyson(chaotic)}
\end{equation}%
where the microvariable $x_{A}^{\prime }$ represents the spacing of the
energy levels and the macrovariable $\mu _{A}^{\prime }$ is the average
spacing. The chain is immersed in the \textit{tilted} magnetic vector field
which has two components $B_{\perp }$ and $B_{\parallel }$ in the
Hamiltonian $\mathcal{H}_{\text{chaotic}}\left( 1,1\right) $. The Gaussian
distribution is identified by information theory as the maximum entropy
distribution if only the expectation value and the variance are known. We
translate this information in the IGAC formalism by coupling the probability
(\ref{wigner-dyson(chaotic)}) to a Gaussian $p_{B}^{\left( \text{Gaussian}%
\right) }\left( x_{B}^{\prime }|\mu _{B}^{\prime },\sigma _{B}^{\prime
}\right) $%
\begin{equation}
p_{B}^{\left( \text{Gaussian}\right) }\left( x_{B}^{\prime }|\mu
_{B}^{\prime },\sigma _{B}^{\prime }\right) =\frac{1}{\sqrt{2\pi \sigma
_{B}^{\prime 2}}}\exp \left( -\frac{\left( x_{B}^{\prime }-\mu _{B}^{\prime
}\right) ^{2}}{2\sigma _{B}^{\prime 2}}\right) ,
\end{equation}%
where the microvariable $x_{B}^{\prime }$ is the intensity of the magnetic
field, the macrovariable $\mu _{B\text{ }}^{\prime }$is the average
intensity of the magnetic energy arising from the interaction of the \textit{%
tilted} magnetic field with the spin $\frac{1}{2}$\ particle magnetic moment
and $\sigma _{B}^{\prime }$ is its covariance:\ during the transition from
the integrable to the chaotic regime, the magnetic field intensity is being
varied (experimentally). It is being tilted and its two components ($%
B_{\perp }$ and $B_{\parallel }$) are being varied as well. Our best guess
based on the experimental mechanism that drives the transitions between the
two regimes is that the magnetic field intensity (actually the microvariable 
$\mu B\cos \varphi $) is Gaussian-distributed (two macrovariables) during
this change. In the chaotic regime, we say the magnetic field intensity is
set to a well-defined value $\left\langle x_{B}^{\prime }\right\rangle =\mu
_{B}^{\prime }$\ with covariance $\sigma _{B}^{\prime }=\sqrt{\left\langle
\left( x_{B}^{\prime }-\left\langle x_{B}^{\prime }\right\rangle \right)
^{2}\right\rangle }$. Thus, the chosen composite probability distribution $%
P^{\left( \text{chaotic}\right) }\left( x_{A}^{\prime },x_{B}^{\prime }|\mu
_{A}^{\prime },\mu _{B}^{\prime },\sigma _{B}^{\prime }\right) $ encoding
relevant information about the system is given by%
\begin{equation}
P^{\left( \text{chaotic}\right) }\left( x_{A}^{\prime },x_{B}^{\prime }|\mu
_{A}^{\prime },\mu _{B}^{\prime },\sigma _{B}^{\prime }\right) =\frac{\pi
\left( 2\pi \sigma _{B}^{\prime 2}\right) ^{-\frac{1}{2}}}{2\mu _{A}^{\prime
2}}x_{A}^{\prime }\exp \left[ -\left( \frac{\pi x_{A}^{\prime 2}}{4\mu
_{A}^{\prime 2}}+\frac{\left( x_{B}^{\prime }-\mu _{B}^{\prime }\right) ^{2}%
}{2\sigma _{B}^{\prime 2}}\right) \right] .
\end{equation}%
Let us denote by $\mathcal{M}_{S}^{\left( \text{chaotic}\right) }$ the
three-dimensional curved statistical manifold underlying our ED model. The
corresponding line element $ds_{\text{chaotic}}^{2}$ on $\mathcal{M}%
_{S}^{\left( \text{chaotic}\right) }$ is given by%
\begin{equation}
ds_{\text{chaotic}}^{2}=ds_{\text{Wigner-Dyson}}^{2}+ds_{\text{Gaussian}%
}^{2}=\frac{4}{\mu _{A}^{\prime 2}}d\mu _{A}^{\prime 2}+\frac{1}{\sigma
_{B}^{\prime 2}}d\mu _{B}^{\prime 2}+\frac{2}{\sigma _{B}^{\prime 2}}d\sigma
_{B}^{\prime 2}.  \label{chaotic case}
\end{equation}%
Applying the IGAC machinery to the line element in (\ref{chaotic case}), we
obtain exponential growth for $\mathcal{V}_{\mathcal{M}_{s}}^{\text{chaotic}%
} $ and linear IGE growth \cite{ca-3, ca-7}, 
\begin{equation}
\mathcal{C}_{\mathcal{M}_{s}}^{\left( \text{chaotic}\right) }\left( \tau
\right) \overset{\tau \rightarrow \infty }{\propto }C_{\text{IG}}\exp \left( 
\mathcal{K}_{\text{IG}}\tau \right) ,\text{ }\mathcal{S}_{\mathcal{M}%
_{s}}^{\left( \text{chaotic}\right) }\left( \tau \right) \overset{\tau
\rightarrow \infty }{\propto }\mathcal{K}_{\text{IG}}\tau .
\label{exponential}
\end{equation}%
The constant\textbf{\ }$C_{\text{IG}}$\textbf{\ }encodes information about
the initial conditions of the statistical macrovariables parametrizing
elements of $\mathcal{M}_{S}^{\left( \text{chaotic}\right) }$. The constant $%
\mathcal{K}_{\text{IG}}$, given by%
\begin{equation}
\mathcal{K}_{\text{IG}}\overset{\tau \rightarrow \infty }{\approx }\frac{d%
\mathcal{S}_{\mathcal{M}_{s}}\left( \tau \right) }{d\tau }\overset{\tau
\rightarrow \infty }{\approx }\lim_{\tau \rightarrow \infty }\left[ \frac{1}{%
\tau }\ln \left\vert \frac{\mathcal{J}_{\mathcal{M}_{S}}\left( \tau \right) 
}{\mathcal{J}_{\mathcal{M}_{S}}\left( 0\right) }\right\vert \right] =\lambda
_{\mathcal{M}_{S}}
\end{equation}%
is the model parameter of the chaotic system and depends on the temporal
evolution of the statistical macrovariables. It plays the role of the
standard Lyapunov exponent of a trajectory and is, in principle, an
experimentally observable quantity. The quantity\textbf{\ }$\mathcal{J}_{%
\mathcal{M}_{S}}\left( \tau \right) $\ is the Jacobi field intensity and%
\textbf{\ }$\lambda _{\mathcal{M}_{S}}$\textbf{\ }may be considered the
information-geometric analogue of the leading Lyapunov exponent in
conventional Hamiltonian systems. Given an explicit expression of\textbf{\ }$%
\mathcal{K}_{\text{IG}}$\textbf{\ }in terms of the observables $\mu
_{A}^{\prime }$, $\mu _{B}^{\prime }$ and $\sigma _{B}^{\prime }$, a clear
understanding of the relation between the IGE (or\textbf{\ }$\mathcal{K}_{%
\text{IG}}$) and the entanglement entropy (or\textbf{\ }$\mathcal{K}_{q}$)
becomes the key point that deserves further study.\textbf{\ }Equations (\ref%
{exponential}) are the information-geometric analogue of the computational
complexity $D_{\varepsilon }^{\left( \text{chaotic}\right) }\left( \tau
\right) $ and the entanglement entropy $\mathcal{S}_{\text{chaotic}}\left(
1,1\right) $ defined in standard quantum information theory, respectively.
This result requires a deeper analysis in order to be fully understood. One
of the major limitations of our findings is the lack of a detailed account
of the comparison of theory with experiment. This point will be among our
primary concerns in future works. Some considerations may however, be
carried out at the present stage. The experimental observables in our
theoretical models are the statistical macrovariables characterizing the
composite probability distributions. In the integrable case, where the
coupling between a Poisson and an Exponential distribution is considered,%
\textbf{\ }$\mu _{A}$ and\textbf{\ }$\mu _{B}$ are the experimental
observables. In the chaotic case, where the coupling between a Wigner-Dyson
and a Gaussian distribution is considered,\textbf{\ }$\mu _{A}^{\prime }$, $%
\mu _{B}^{\prime }$ and $\sigma _{B}^{\prime }$\textbf{\ }play the role of
the experimental observables. We think one way to test our theory may be to
determine a numerical estimate of the leading Lyapunov exponent\textbf{\ }$%
\lambda _{\text{max}}$\textbf{\ }or the Lyapunov spectrum for the
Hamiltonian systems under investigation directly from experimental data
(measurement of a time series) and compare it to our theoretical estimate for%
\textbf{\ }$\lambda _{\mathcal{M}_{S}}$\textbf{\ }\cite{wolf}.\textbf{\ }%
However, we are aware that it may be rather difficult to evaluate Lyapunov
exponents numerically. Otherwise, knowing that the mean values of the
positive Lyapunov exponents are related to the Kolmogorov-Sinai (KS)
dynamical entropy, we suggest to measure the KS entropy\textbf{\ }$\mathcal{K%
}$\textbf{\ }directly from a time signal associated with a suitable
combination of our experimental observables and compare it to our indirect
theoretical estimate for\textbf{\ }$\mathcal{K}_{\text{IG}}$\textbf{\ }from
the asymptotic behaviors of our statistical macrovariables \cite{procaccia}.
We are aware that the basis of our discussion is rather qualitative.
However, we hope that with additional study, especially in clarifying the
relation between the IGE and the entanglement entropy, our theoretical
information geometric characterization will find experimental support in the
future. For these reasons\textbf{,} the statement that our findings may be
relevant to experiments verifying the existence of chaoticity and related
dynamical properties on a macroscopic level in energy level statistics in
chaotic and regular quantum spin chains is purely a conjecture at this stage.

\subsection{Quantum Entangled Wave Packets}

As our final example, we apply the IGAC to characterize the quantum
entanglement produced by a head-on collision between two identical (but
distinguishable) spinless, structureless, non-relativistic particles of mass 
$m$, each represented by minimum uncertainty Gaussian wave-packets
interacting via a scattering process\ \cite{Wang}. Before colliding, the two
particles are in the form of disentangled Gaussian wave packets, each
characterized by a width $\sigma _{\mathrm{o}}$ in momentum space. The
initial distance between the two particles is $R_{\mathrm{o}}$ and their
average initial momenta - setting the Planck constant $\hbar $ equal to one
- are $\mp k_{\mathrm{o}}$, respectively. After some straightforward algebra 
\cite{Wang} it can be shown that the initial (pre-collisional) two-particle
square wave amplitude in momentum space is given by%
\begin{equation}
P_{\text{pre}}^{\text{(QM)}}\left( k_{1},k_{2}|k\text{$_{\mathrm{o}}$}%
,\sigma _{\mathrm{o}}\right) =\frac{1}{2\pi \sigma _{k_{\mathrm{o}}}^{2}}%
\exp \left[ -\frac{\left( k_{1}-k_{\mathrm{o}}\right) ^{2}+\left( k_{2}+k_{%
\mathrm{o}}\right) ^{2}}{2\sigma _{k_{\mathrm{o}}}^{2}}\right] \text{,}
\label{preqm}
\end{equation}%
where $\sigma _{k_{\mathrm{o}}}=\frac{\sigma _{\mathrm{o}}}{\hbar }$, $\pm
k_{\mathrm{o}}=\left\langle k_{1/2}\right\rangle _{\mathrm{o}}=\frac{%
\left\langle p_{1/2}\right\rangle _{\mathrm{o}}}{\hbar }=\pm \frac{p_{%
\mathrm{o}}}{\hbar }$ with $\left\langle \mathbf{p}_{1}\right\rangle _{%
\mathrm{o}}=\mathbf{p}_{\mathrm{o}}$, $\left\langle \mathbf{p}%
_{2}\right\rangle _{\mathrm{o}}=-\mathbf{p}_{\mathrm{o}}$, $\sigma _{\mathrm{%
o}}$ is defined as in (\ref{var}) and we have made use of the center of mass
and relative coordinates whose conjugate momenta are defined as $K\equiv
k_{1}+k_{2}\in \left( -\infty ,+\infty \right) $ and $k\equiv \frac{1}{2}%
\left( k_{1}-k_{2}\right) \in \left( -\infty ,+\infty \right) $ with $%
k_{1/2}=\frac{p_{1/2}}{\hbar }\in \left( -\infty ,+\infty \right) $.

Similarly, following \cite{Wang}, and after some tedious algebra, one finds
that the final (long time limit, post-collisional) two-particle square wave
amplitude in momentum space is given by%
\begin{equation}
P_{\text{post}}^{\text{(QM)}}\left( k_{1},k_{2}|k_{\mathrm{o}},\sigma _{k_{%
\mathrm{o}}};r_{\mathrm{QM}}\right) \simeq \frac{\exp \left\{ -\frac{1}{%
2\left( 1-r_{\mathrm{QM}}^{2}\right) }\left[ \frac{\left( k_{1}-k_{\mathrm{o}%
}\right) ^{2}}{\sigma _{k_{\mathrm{o}}}^{2}}-2r_{\mathrm{QM}}\frac{\left(
k_{1}-k_{\mathrm{o}}\right) \left( k_{2}+k_{\mathrm{o}}\right) }{\sigma _{k_{%
\mathrm{o}}}^{2}}+\frac{\left( k_{2}+k_{\mathrm{o}}\right) ^{2}}{\sigma _{k_{%
\mathrm{o}}}^{2}}\right] \right\} }{2\pi \sigma _{k_{\mathrm{o}}}^{2}\sqrt{%
1-r_{\mathrm{QM}}^{2}}},  \label{postqm}
\end{equation}%
with%
\begin{equation}
r_{\mathrm{QM}}\equiv \sqrt{8\left( 2k_{\mathrm{o}}^{2}+\sigma _{k_{\mathrm{o%
}}}^{2}\right) R_{\mathrm{o}}a_{\mathrm{s}}}\ll 1,  \label{rqm}
\end{equation}%
where the parameter $a_{\mathrm{s}}$ has dimension of length and is defined
as the $s$-wave scattering length \cite{Landau}. It is evident from (\ref%
{rqm}) that $r_{\mathrm{QM}}$ is non-zero and positive. Thus, $f\left( k_{%
\mathrm{o}}\right) =\frac{e^{i\theta _{\mathrm{o}}}\sin \theta _{\mathrm{o}}%
}{k_{\mathrm{o}}}=\overset{\theta \left( k_{\mathrm{o}}\right) \ll 1}{%
\approx }\frac{\theta \left( k_{\mathrm{o}}\right) }{k_{\mathrm{o}}}+%
\mathcal{O}\left( \theta ^{2}\right) \overset{k_{\mathrm{o}}L\ll 1}{\approx }%
-a_{\mathrm{s}}$, where $\theta _{\mathrm{o}}\equiv \theta \left( k_{\mathrm{%
o}}\right) \approx -\frac{2\eta Vk_{\mathrm{o}}L^{3}}{3\hbar ^{2}}\approx
-k_{\mathrm{o}}a_{\mathrm{s}}$ denotes the $s$-wave scattering phase shift, $%
\mu $ is the reduced mass $\mu =m/2$, $f\left( k\right) $ is the scattering
amplitude and $L$ is the range of the scattering potential\ $V$ given by 
\begin{equation}
V(x)=\left\{ 
\begin{array}{ll}
V, & \;0\leq x\leq L \\ 
0, & \;x>L%
\end{array}%
\right. ,
\end{equation}%
where $V$ denotes the height (for $V>0$; repulsive potential) or depth (for $%
V<0$; attractive potential) of the potential. The quantity $\theta \left(
k\right) $ is the $s$-wave scattering phase shift considered around $k=k_{%
\mathrm{o}}$ (i.e. assuming our wave-packet is well-localized around $k=k_{%
\mathrm{o}}$) and in the limit of low-energy scattering, i.e. $\theta \left(
k\right) \ll 1$.

We conjecture that the quantum entanglement produced by a head-on collision
between two Gaussian wave packets are macroscopic manifestations emerging
from specific underlying microscopic statistical structures. Specifically,
we propose that $P_{\text{pre}}^{\text{(QM)}}\left( k_{1},k_{2}|k_{\mathrm{o}%
},\sigma _{k_{\mathrm{o}}}\right) $ can be interpreted as a limiting case
(initial time limit) arising from a Gaussian probability distribution $P_{%
\text{pre}}^{\text{(IG)}}=\left( x,y|\mu _{x},\mu _{y};\sigma \right) $ ,%
\begin{equation}
P_{\text{pre}}^{\text{(IG)}}\left( x,y|\mu _{x},\mu _{y},\sigma \right) 
\overset{\text{def}}{=}\frac{1}{2\pi \sigma ^{2}}\exp \left[ -\frac{\left(
x-\mu _{x}\right) ^{2}}{2\sigma ^{2}}-\frac{\left( y-\mu _{y}\right) ^{2}}{%
2\sigma ^{2}}\right] .  \label{igpre1}
\end{equation}%
Upon setting $x\rightarrow k_{1}$, $y\rightarrow k_{2}$, $\left\langle
x\right\rangle =\mu _{x}\rightarrow \mu _{k_{1}}\equiv +k_{\mathrm{o}}$, $%
\left\langle y\right\rangle =\mu _{y}\rightarrow \mu _{k_{2}}\equiv -k_{%
\mathrm{o}}$ and $\sigma \rightarrow \sigma _{k_{\mathrm{o}}}$, we obtain $%
P_{\text{pre}}^{\text{(IG)}}=\left( x,y|\mu _{x},\mu _{y};\sigma \right)
\rightarrow P_{\text{pre}}^{\text{(IG)}}\left( k_{1},k_{2}|\mu _{k_{1}},\mu
_{k_{2}},\sigma \right) $%
\begin{equation}
P_{\text{pre}}^{\text{(IG)}}\left( k_{1},k_{2}|\mu _{k_{1}},\mu
_{k_{2}},\sigma \right) \overset{\text{def}}{=}\frac{1}{2\pi \sigma ^{2}}%
\exp \left[ -\frac{\left( k_{1}-\mu _{k_{1}}\right) ^{2}}{2\sigma ^{2}}-%
\frac{\left( k_{2}-\mu _{k_{2}}\right) ^{2}}{2\sigma ^{2}}\right]
\label{igpre}
\end{equation}%
enabling the identification%
\begin{equation}
P_{\text{pre}}^{\text{(QM)}}\left( k_{1},k_{2}|k_{\mathrm{o}},\sigma _{k_{%
\mathrm{o}}}\right) =P_{\text{pre}}^{\text{(IG)}}\left( k_{1},k_{2}|k_{%
\mathrm{o}},\sigma _{k_{\mathrm{o}}}\right) .  \label{A}
\end{equation}%
The variances $\sigma _{x}$ and $\sigma _{y}$ in the random variables $x$
and $y$ respectively, are given by the standard definition (\ref{var}). We
remark that in general, $\sigma _{x}\neq \sigma _{y}$. In the present
example however, it is sufficient to consider $\sigma _{x}=\sigma
_{y}=\sigma $.

We propose that $P_{\text{post}}^{\text{(QM)}}\left( k_{1},k_{2}|k_{\mathrm{o%
}},\sigma _{k_{\mathrm{o}}};r_{\mathrm{QM}}\right) $ can be viewed as a
limiting case (final or long time limit) arising from a Gaussian probability
distribution $P_{\text{post}}^{\text{(IG)}}\left( x,y|\mu _{x},\mu
_{y};\sigma ,r\right) $,%
\begin{equation}
P_{\text{post}}^{\text{(IG)}}\left( x,y|\mu _{x},\mu _{y};\sigma ,r\right) =%
\frac{\exp \left\{ -\frac{1}{2\left( 1-r^{2}\right) }\left[ \frac{\left(
x-\mu _{x}\right) ^{2}}{\sigma ^{2}}-2r\frac{\left( x-\mu _{x}\right) \left(
y-\mu _{y}\right) }{\sigma ^{2}}+\frac{\left( y-\mu _{y}\right) ^{2}}{\sigma
^{2}}\right] \right\} }{2\pi \sigma ^{2}\sqrt{1-r^{2}}},  \label{igpost}
\end{equation}%
where the micro-correlation coefficient $r$ is defined as in (\ref{microC}).
In the present example, the micro-correlation coefficient $r$ is considered
to have compact support over the line segment $[0,1)$, that is $r\in \lbrack
0,1)$. Upon setting $\mu _{k_{1}}\rightarrow +k_{\mathrm{o}}$, $\mu
_{k_{2}}\rightarrow -k_{\mathrm{o}}$ and $\sigma \rightarrow \sigma _{k_{%
\mathrm{o}}}$ we obtain%
\begin{equation}
P_{\text{post}}^{\text{(IG)}}\left( k_{1},k_{2}|k_{\mathrm{o}},\sigma _{k_{%
\mathrm{o}}};r\right) =\frac{\exp \left\{ -\frac{1}{2\left( 1-r^{2}\right) }%
\left[ \frac{\left( k_{1}-k_{\mathrm{o}}\right) ^{2}}{\sigma _{k_{\mathrm{o}%
}}^{2}}-2r\frac{\left( k_{1}-k_{\mathrm{o}}\right) \left( k_{2}+k_{\mathrm{o}%
}\right) }{\sigma _{k_{\mathrm{o}}}^{2}}+\frac{\left( k_{2}+k_{\mathrm{o}%
}\right) ^{2}}{\sigma _{k_{\mathrm{o}}}^{2}}\right] \right\} }{2\pi \sigma
_{k_{\mathrm{o}}}^{2}\sqrt{1-r^{2}}}.  \label{postig}
\end{equation}%
In this case, when both the weak correlation ($r\ll 1$) and the weak
scattering conditions ($\left\vert \theta \left( k_{\mathrm{o}}\right)
\right\vert \ll 1$) are satisfied, we obtain an excellent overlapping
between (\ref{postqm}) and (\ref{postig}), so that%
\begin{equation}
P_{\text{post}}^{\text{(QM)}}\left( k_{1},k_{2}|k_{\mathrm{o}},\sigma _{k_{%
\mathrm{o}}};r_{\mathrm{QM}}\right) \simeq P_{\text{post}}^{\text{(IG)}%
}\left( k,k_{\mathrm{o}},\sigma _{k_{\mathrm{o}}};r\right) \text{ for }r\ll 1%
\text{, }r_{\mathrm{QM}}\ll 1\text{ and }\left\vert \theta \left( k_{\mathrm{%
o}}\right) \right\vert \ll 1,  \label{B}
\end{equation}%
assuming that $k_{\mathrm{o}}$, $\sigma _{k_{\mathrm{o}}}$, $r$ and $r_{%
\mathrm{QM}}$ are fixed numerical constants, and letting $k_{1/2}$ assume
values in the neighborhood of $k_{\mathrm{o}}$.

At this stage our conjecture is only mathematically sustained by the formal
identities (\ref{A}) and (\ref{B}). To render our conjecture physically
relevant, recall that $s$-wave scattering can also be described in terms of
a scattering potential $V(x)$ and the scattering phase shift $\theta \left(
k\right) $. Integrating the radial part of the Schr\"{o}dinger equation with
this potential for the scattered wave and imposing the matching condition at 
$x=L$ for its solution and its first derivative leads to \cite{Hunt}%
\begin{equation}
k_{\mathrm{in}}\cot \left( k_{\mathrm{in}}L\right) =k_{\mathrm{out}}\cot
\left( k_{\mathrm{out}}L+\theta \right) \text{,}  \label{match}
\end{equation}%
with%
\begin{eqnarray}
k_{\mathrm{in}} &=&\frac{\sqrt{2\mu \left( \mathcal{E}-V\right) }}{\hbar }%
,\;\;\;0<x<L,  \label{eq:ka} \\
k_{\mathrm{out}} &=&\frac{\sqrt{2\mu \mathcal{E}}}{\hbar },\;\;\;x>L,
\label{eq:kb}
\end{eqnarray}%
The quantities $\mu $ and $\mathcal{E}$ are the reduced mass and kinetic
energy of the two-particle system in the relative coordinates, respectively; 
$k_{\mathrm{in}}$ and $k_{\mathrm{out}}$ represent\textbf{\ }the
conjugate-coordinate wave vectors inside and outside the potential region,
respectively. Equation (\ref{match}) indicates that the scattering potential 
$V(x)$ shifts the phase of the scattered wave at points beyond the
scattering region.

Our information geometric modeling may be briefly described in the following
way. The pre-collisional scenario is characterized by the information
geometric dynamics on the curved statistical manifold $\mathcal{M}%
_{s}^{\left( \text{uncorr.}\right) }$ of uncorrelated Gaussian probability
distributions $P_{\text{pre}}^{\text{(IG)}}\left( k_{1},k_{2}|\mu
_{k_{1}},\mu _{k_{2}},\sigma \right) $ given in (\ref{igpre}). The geodesic
trajectories on $\mathcal{M}_{s}^{\left( \text{uncorr.}\right) }$ for the
non-correlated Gaussian system is given by%
\begin{eqnarray}
\left\langle p_{1\mathrm{b}}(\tau )\right\rangle &=&\mu _{1}\left( \tau
;0\right) =-\sqrt{p_{\mathrm{o}}^{2}+2\sigma _{\mathrm{o}}^{2}}\tanh \left(
A_{\mathrm{o}}\tau \right) ,  \label{eq:p1b} \\
\left\langle p_{2\mathrm{b}}(\tau )\right\rangle &=&\mu _{2}\left( \tau
;0\right) =\sqrt{p_{\mathrm{o}}^{2}+2\sigma _{\mathrm{o}}^{2}}\tanh \left(
A_{\mathrm{o}}\tau \right) ,  \label{eq:p2b} \\
\left\langle \sigma _{\mathrm{b}}(\tau )\right\rangle &=&\sigma \left( \tau
;0\right) =\sqrt{\frac{1}{2}p_{\mathrm{o}}^{2}+\sigma _{\mathrm{o}}^{2}}%
\frac{1}{\cosh \left( A_{\mathrm{o}}\tau \right) }.  \label{eq:sigmab}
\end{eqnarray}%
The post-collisional scenario is characterized by the information geometric
dynamics on the curved statistical manifold $\mathcal{M}_{s}^{\left( \text{%
corr.}\right) }$ of correlated Gaussian probability distributions $P_{\text{%
post}}^{\text{(IG)}}\left( k_{1},k_{2}|k_{\mathrm{o}},\sigma _{k_{\mathrm{o}%
}};r\right) $ given in (\ref{postig}). The geodesic trajectories on $%
\mathcal{M}_{s}^{\left( \text{corr.}\right) }$ for the non-correlated
Gaussian system is given by%
\begin{eqnarray}
\left\langle p_{1\mathrm{a}}(\tau )\right\rangle &=&\mu _{1}(\tau ;r)=-\sqrt{%
\left( 1-r\right) \left( p_{\mathrm{o}}^{2}+2\sigma _{\mathrm{o}}^{2}\right) 
}\tanh \left( A_{\mathrm{o}}\tau \right) ,  \label{eq:p1a} \\
\left\langle p_{2\mathrm{a}}(\tau )\right\rangle &=&\mu _{2}(\tau ;r)=\sqrt{%
\left( 1-r\right) \left( p_{\mathrm{o}}^{2}+2\sigma _{\mathrm{o}}^{2}\right) 
}\tanh \left( A_{\mathrm{o}}\tau \right) ,  \label{eq:p2a} \\
\left\langle \sigma _{\mathrm{a}}(\tau )\right\rangle &=&\sigma (\tau ;r)=%
\sqrt{\frac{1}{2}p_{\mathrm{o}}^{2}+\sigma _{\mathrm{o}}^{2}}\frac{1}{\cosh
\left( A_{\mathrm{o}}\tau \right) },  \label{eq:sigmaa}
\end{eqnarray}%
where the subscript\textbf{\ }\textquotedblleft \textbf{$_{\mathrm{o}}$}%
\textquotedblright\ denotes the initial state, the subscripts
\textquotedblleft $_{1}$\textquotedblright\ and \textquotedblleft $_{2}$%
\textquotedblright ${}$ denote particle $1$ and particle $2$, respectively;
subscripts \textquotedblleft $_{\mathrm{b}}$\textquotedblright ${}$ and
\textquotedblleft $_{\mathrm{a}}$\textquotedblright ${}$ denote `before' and
`after' collision, respectively and 
\begin{eqnarray}
A_{\mathrm{o}} &\equiv &\frac{1}{\tau _{\mathrm{o}}}\sinh ^{-1}\left( \frac{%
p_{\mathrm{o}}}{\sqrt{2}\sigma _{\mathrm{o}}}\right)  \label{a0} \\
&\overset{\frac{\sigma _{\mathrm{o}}}{p_{\mathrm{o}}}\ll 1}{=}&\frac{1}{\tau
_{\mathrm{o}}}\left\{ \ln \left( \frac{\sqrt{2}p_{\mathrm{o}}}{\sigma _{%
\mathrm{o}}}\right) +\frac{1}{2}\left( \frac{\sigma _{\mathrm{o}}}{p_{%
\mathrm{o}}}\right) ^{2}-\frac{3}{8}\left( \frac{\sigma _{\mathrm{o}}}{p_{%
\mathrm{o}}}\right) ^{4}+\mathcal{O}\left[ \left( \frac{\sigma _{\mathrm{o}}%
}{p_{\mathrm{o}}}\right) ^{6}\right] \right\} .  \notag
\end{eqnarray}%
The two sets of geodesic curves comprised of $\left\{ \left\langle p_{1%
\mathrm{b}}(\tau )\right\rangle ,\,\left\langle p_{2\mathrm{b}}(\tau
)\right\rangle ,\,\left\langle \sigma _{\mathrm{b}}(\tau )\right\rangle
\right\} $ (for the non-correlated model) and $\left\{ \left\langle p_{1%
\mathrm{a}}(\tau )\right\rangle ,\,\left\langle p_{2\mathrm{a}}(\tau
)\right\rangle ,\,\left\langle \sigma _{\mathrm{a}}(\tau )\right\rangle
\right\} $ (for the correlated model) are joined at the junction $\tau =0$: $%
\tau <0$ (before collision) for the non-correlated model and $\tau \geq 0$
(after collision) for the correlated model. We recognize that the momenta\ $%
\left\langle p_{1\mathrm{b}}(\tau )\right\rangle $ and\ $\left\langle p_{1%
\mathrm{a}}(\tau )\right\rangle $ asymptotically converge to\ $\sqrt{p_{%
\mathrm{o}}^{2}+2\sigma _{\mathrm{o}}^{2}}$\ and $-\sqrt{\left( 1-r\right)
\left( p_{\mathrm{o}}^{2}+2\sigma _{\mathrm{o}}^{2}\right) }$\ toward $\tau
=-\infty $ and $\tau =+\infty $, respectively (the same is true for\ $%
-\left\langle p_{2\mathrm{b}}(\tau )\right\rangle $ and\ $-\left\langle p_{2%
\mathrm{a}}(\tau )\right\rangle $) while $\left\langle \sigma _{\mathrm{b}%
}(\tau )\right\rangle $ and $\left\langle \sigma _{\mathrm{a}}(\tau
)\right\rangle $ are identical and vanishingly small toward $\tau =\pm
\infty $. Furthermore, we observe that there is continuity between $%
\left\langle p_{1/2\mathrm{b}}(\tau )\right\rangle $ and $\left\langle p_{1/2%
\mathrm{a}}(\tau )\right\rangle $ and between $\left\langle \sigma _{\mathrm{%
b}}(\tau )\right\rangle $ and $\left\langle \sigma _{\mathrm{a}}(\tau
)\right\rangle $ at the junction, $\tau =0$

A question that now arises is how to determine the scattering phase shift in
view of the fact that our statistical model is correlated after collision.
Initially, we need to examine how correlations affect the momentum geodesic
curve $\left\langle p_{1/2}(\tau )\right\rangle $. For this purpose we
define the\textbf{\ }momentum-difference curve $\left\langle p(\tau
)\right\rangle \equiv \frac{1}{2}\left[ \left\langle p_{2}(\tau
)\right\rangle -\left\langle p_{1}(\tau )\right\rangle \right] $. Comparison
of the following two equations 
\begin{eqnarray}
\left\langle p\left( \tau ;0\right) \right\rangle &\equiv &\frac{1}{2}\left[
\left\langle p_{2\mathrm{b}}(\tau )\right\rangle -\left\langle p_{1\mathrm{b}%
}(\tau )\right\rangle \right] =\sqrt{p_{\mathrm{o}}^{2}+2\sigma _{\mathrm{o}%
}^{2}}\tanh \left( A_{\mathrm{o}}\tau \right) ,  \label{eq:p_0} \\
\left\langle p\left( \tau ;r\right) \right\rangle &\equiv &\frac{1}{2}\left[
\left\langle p_{2\mathrm{a}}(\tau )\right\rangle -\left\langle p_{1\mathrm{a}%
}(\tau )\right\rangle \right] =\sqrt{\left( 1-r\right) \left( p_{\mathrm{o}%
}^{2}+2\sigma _{\mathrm{o}}^{2}\right) }\tanh \left( A_{\mathrm{o}}\tau
\right) ,  \label{eq:p_r}
\end{eqnarray}%
which follow from (\ref{eq:p1b}), (\ref{eq:p2b}), (\ref{eq:p1a}) and (\ref%
{eq:p2a}), indicates that at any arbitrary time $\tau \geq 0$%
\begin{equation}
\left\langle p\left( \tau ;0\right) \right\rangle \geq \left\langle p\left(
\tau ;r\right) \right\rangle ,  \label{eq:p_compare}
\end{equation}%
while both (\ref{eq:p_0}) and (\ref{eq:p_r}) share the functional argument $%
A_{\mathrm{o}}\tau $. Condition (\ref{eq:p_compare}) implies that the
correlation causes a reduction in the momentum for any $\tau \geq 0$
(relative to the non-correlated case)\textbf{.} This situation is analogous
to the change in momentum caused by a repulsive scattering potential (see (%
\ref{eq:ka}) and (\ref{eq:kb})). It is then reasonable to assume there
exists some connection between the scattering potential and the correlation.
Provided this connection is established, one should be able to determine the
scattering phase shift in terms of the correlation via equations (\ref{match}%
), (\ref{eq:ka}) and (\ref{eq:kb}). In this way, one can ultimately
establish a connection between quantum entanglement and the statistical
micro-correlation.

Recall that before collision (at the affine time $-\tau _{\mathrm{o}}$)
particles\ $1$\ and\ $2$\ are separated by\ a linear distance\ $R_{\mathrm{o}%
}$. Each particle has momenta\ $p_{\mathrm{o}}$ and\ $-p_{\mathrm{o}}$,
respectively and the same momentum spread\ $\sigma _{\mathrm{o}}$. Then from
(\ref{eq:p1b}), (\ref{eq:p2b}) and (\ref{eq:sigmab}) we have%
\begin{eqnarray}
p_{\mathrm{o}} &=&\left\langle p_{1\mathrm{b}}\left( -\tau _{\mathrm{o}%
}\right) \right\rangle =-\left\langle p_{2\mathrm{b}}\left( -\tau _{\mathrm{o%
}}\right) \right\rangle =\sqrt{p_{\mathrm{o}}^{2}+2\sigma _{\mathrm{o}}^{2}}%
\tanh \left( A_{\mathrm{o}}\tau _{\mathrm{o}}\right) ,  \label{eq:pb0} \\
\sigma _{\mathrm{o}} &=&\left\langle \sigma _{\mathrm{b}}\left( -\tau _{%
\mathrm{o}}\right) \right\rangle =\sqrt{\frac{1}{2}p_{\mathrm{o}}^{2}+\sigma
_{\mathrm{o}}^{2}}\frac{1}{\cosh \left( A_{\mathrm{o}}\tau _{\mathrm{o}%
}\right) }.  \label{eq:sigmab0}
\end{eqnarray}

For arbitrary $\tau \geq 0$ after collision, the system of particles $1$ and 
$2$, which initially carried momenta $p_{\mathrm{o}}$ and $-p_{\mathrm{o}}$,
respectively at $\tau =-\tau _{\mathrm{o}}$ before collision, now carries
the relative conjugate-momentum $\left\langle p\left( \tau \text{; }r\right)
\right\rangle $ given by (\ref{eq:p_r}) due to the correlation. With
nonvanishing micro-correlation the wave-packets experience the effect of a
repulsive potential; the magnitude of the wave vectors (or momenta)
decreases relative to the corresponding non-correlated value. One may
rewrite (\ref{match}), (\ref{eq:ka}) and (\ref{eq:kb}) as%
\begin{equation}
k_{r}\cot \left( k_{r}L\right) =k_{\mathrm{o}}\cot \left( k_{\mathrm{o}%
}L+\theta _{\mathrm{o}}\right) ,  \label{eq:cot_r0}
\end{equation}%
with%
\begin{eqnarray}
k_{r} &=&\frac{\sqrt{2\mu \left( \mathcal{E}-V\right) }}{\hbar },\;0<x<L,
\label{eq:k_r} \\
k_{\mathrm{o}} &=&\frac{\sqrt{2\mu \mathcal{E}}}{\hbar },\;x>L,
\label{eq:k_0}
\end{eqnarray}%
where $k_{r}$ and $k_{\mathrm{o}}$ represent the wave vectors with and
without the correlation, respectively. The connection between the
correlation and the scattering potential can be established by combining (%
\ref{eq:k_r}) and (\ref{eq:k_0}). From (\ref{eq:p_compare}) one finds that
the correlation renders 
\begin{equation}
k_{\mathrm{o}}\text{ }\longrightarrow \text{ }k_{r}\equiv \sqrt{1-r}k_{%
\mathrm{o}}.  \label{eq:k_reduce}
\end{equation}%
Then using (\ref{eq:k_r}), (\ref{eq:k_0}) and (\ref{eq:k_reduce}), we
determine the scattering potential, 
\begin{equation}
V=r\mathcal{E}=r\frac{\hbar ^{2}k_{\mathrm{o}}^{2}}{2\mu }=r\frac{p_{\mathrm{%
o}}^{2}}{2\mu }.  \label{eq:potential}
\end{equation}%
Equation (\ref{eq:potential}) clearly establishes a connection between the
correlation coefficient and the scattering potential: the correlation
coefficient is the ratio of the scattering potential to the initial relative
kinetic energy of the system. From (\ref{eq:potential}) it is evident that
our interaction potential is repulsive, i.e. $V>0$ since we consider
non-negative micro-correlations, $r\in \lbrack 0,1)$.

With the potential determined, one can determine the scattering phase shift $%
\theta _{\mathrm{o}}$ (for low energy $s$-wave scattering)\ by combining
equations (\ref{eq:cot_r0}), (\ref{eq:k_r}), (\ref{eq:k_0}) and (\ref%
{eq:potential}), the result being,%
\begin{equation}
\tan \theta _{\mathrm{o}}\overset{k_{\mathrm{o}}L=p_{\mathrm{o}}L/\hbar \ll 1%
}{\approx }\theta _{\mathrm{o}}=-\frac{r\left( k_{\mathrm{o}}L\right) ^{3}}{3%
}.  \label{eq:phase_shift2}
\end{equation}%
By means of (\ref{eq:potential}) and (\ref{eq:phase_shift2}) we can express
the scattering phase shift in terms of the scattering potential 
\begin{equation}
\theta _{\mathrm{o}}\approx -\frac{2\mu Vk_{\mathrm{o}}L^{3}}{3\hbar ^{2}}=-%
\frac{2\mu Vp_{\mathrm{o}}L^{3}}{3\hbar ^{3}},  \label{eq:phase_shift_V}
\end{equation}%
which is in agreement with \cite{Mishima}. This is the first significant
finding that allows to state that our conjecture is also physically
motivated.

As the scattering potential has been determined, so too can the scattering
amplitude be determined. To this end, we write 
\begin{equation}
f\left( k_{\mathrm{o}}\right) =\frac{e^{i\theta _{\mathrm{o}}}\sin \theta _{%
\mathrm{o}}}{k_{\mathrm{o}}}\approx \frac{\theta _{\mathrm{o}}}{k_{\mathrm{o}%
}}\approx -a_{\mathrm{s}}  \label{eq:f01}
\end{equation}%
for low energy $s$-wave scattering, $k_{\mathrm{o}}L=p_{\mathrm{o}}L/\hbar
\ll 1$. Thus, we finally obtain the scattering cross section: 
\begin{equation}
\Sigma =4\pi \left\vert f\left( k_{\mathrm{o}}\right) \right\vert
^{2}\approx \frac{4\pi r^{2}k_{\mathrm{o}}^{4}L^{6}}{9}=\frac{16\pi \mu
^{2}V^{2}L^{6}}{9\hbar ^{4}}\approx 4\pi a_{\mathrm{s}}^{2}.
\label{eq:cross-section}
\end{equation}%
In order to properly analyze entanglement, the entanglement entropy obtained
from the long time limit post-collisional wave function is required. In most
cases however, this must be performed numerically. Thus, to approach the
problem analytically and simultaneously gain insights into the problem, it
is convenient to make use of the linearized version of the entropy of the
system, i.e. of the purity of the system \cite{Wang}. The purity function is
defined as%
\begin{equation}
\mathcal{P}\overset{\text{def}}{=}\mathrm{Tr}\left( \rho _{A}^{2}\right) ,
\end{equation}%
where $\rho _{A}\equiv \mathrm{Tr}_{B}\left( \rho _{AB}\right) $ is the
reduced density matrix of particle $A$ and $\rho _{AB}$ is the two-particle
density matrix associated with the post-collisional two-particle wave
function. For pure two-particle states, the smaller the value of $\mathcal{P}
$ the higher the entanglement. That is, the loss of purity provides an
indicator of the degree of entanglement. Hence, a disentangled product state
corresponds to $\mathcal{P}=1$. Under the assumption that the two particles
are well separated both initially (before collision) and finally (after
collision), and further assuming that the colliding Gaussian wave packets
are very narrow in momentum space ($\sigma _{k_{\mathrm{o}}}\ll 1$ such that
the phase shift can be treated as a constant $\theta \left( k_{\mathrm{o}%
}\right) $), it follows that the purity of the post-collisional two-particle
wave function is approximately given by \cite{Wang}%
\begin{equation}
\mathcal{P}=1-8\left( 2k_{\mathrm{o}}^{2}+\sigma _{k_{\mathrm{o}%
}}^{2}\right) R_{\mathrm{o}}a_{\mathrm{s}}+\mathcal{O}\left( a_{\mathrm{s}%
}^{2}\right) .  \label{eq:P1}
\end{equation}%
Employing the scattering cross section $\Sigma =4\pi \left\vert f\left( k_{%
\mathrm{o}}\right) \right\vert ^{2}\approx \frac{4\pi r^{2}k_{\mathrm{o}%
}^{4}L^{6}}{9}=\frac{16\pi \mu ^{2}V^{2}L^{6}}{9\hbar ^{4}}\approx 4\pi a_{%
\mathrm{s}}^{2}$, we may express the purity in an alternative manner, namely,%
\begin{equation}
\mathcal{P}=1-\frac{4\left( 2k_{\mathrm{o}}^{2}+\sigma _{k_{\mathrm{o}%
}}^{2}\right) R_{\mathrm{o}}\sqrt{\Sigma }}{\sqrt{\pi }}+\mathcal{O}\left(
\Sigma \right) .  \label{eq:P2}
\end{equation}%
Equations (\ref{eq:P1}) and (\ref{eq:P2}) above demonstrate how the
entanglement can be measured from the loss of purity by use of the
scattering length or cross section. By combining (\ref{eq:P1}) and the
square of (\ref{eq:f01}) we find the purity%
\begin{equation}
\mathcal{P}\approx 1-\frac{8rk_{\mathrm{o}}^{2}\left( 2k_{\mathrm{o}%
}^{2}+\sigma _{k_{\mathrm{o}}}^{2}\right) R_{\mathrm{o}}L^{3}}{3}=1-\frac{%
16\mu V\left( 2k_{\mathrm{o}}^{2}+\sigma _{k_{\mathrm{o}}}^{2}\right) R_{%
\mathrm{o}}L^{3}}{3\hbar ^{2}}.  \label{eq:purity}
\end{equation}%
Equation (\ref{eq:purity}) implies that the purity $\mathcal{P}$ can be
expressed in terms of physical quantities such as the scattering potential
height $V$ and the initial quantities $k_{\mathrm{o}}$, $\sigma _{\mathrm{o}%
} $ and $R_{\mathrm{o}}$ via (\ref{eq:cross-section}) and (\ref{eq:potential}%
). This is the second significant finding obtained within our hybrid
approach (quantum dynamical results combined with information geometric
modeling techniques) that allows to explain how the interaction potential
height $V$ and the incident particle energies$\mathcal{E}$ control the
strength of the entanglement. The role played by $r$ in the quantities $%
\mathcal{P}$ and $V$ suggests that the physical information about quantum
scattering - and therefore about quantum entanglement - is encoded in the
statistical correlation coefficient, specifically in the covariance term $%
\mathrm{Cov}\left( k_{1},k_{2}\right) \overset{\text{def}}{=}\left\langle
k_{1}k_{2}\right\rangle -\left\langle k_{1}\right\rangle \left\langle
k_{2}\right\rangle $ appearing in the definition of $r$.

The correlation coefficient $r$ can now be expressed in terms of the
physical quantities such as the scattering potential, the scattering cross
section and the purity. Solving equations (\ref{eq:potential}), (\ref%
{eq:cross-section}) and (\ref{eq:purity}) for $r$, we obtain%
\begin{eqnarray}
r &=&\frac{V}{\mathcal{E}}=\frac{2\mu V}{\hbar ^{2}k_{\mathrm{o}}^{2}}=\frac{%
2\mu V}{p_{\mathrm{o}}^{2}},  \label{eq:corr_r1} \\
&\approx &\frac{3\sqrt{\Sigma }}{2\sqrt{\pi }k_{\mathrm{o}}^{2}L^{3}},
\label{eq:corr_r2} \\
&\approx &\frac{3\left( 1-\mathcal{P}\right) }{8k_{\mathrm{o}}^{2}\left( 2k_{%
\mathrm{o}}^{2}+\sigma _{k\mathrm{o}}^{2}\right) R_{\mathrm{o}}L^{3}}.
\label{eq:corr_r3}
\end{eqnarray}%
In view of (\ref{rqm}), (\ref{eq:cross-section}) and (\ref{eq:corr_r2}), one
obtains the following relation: 
\begin{equation}
\frac{V}{L^{3}}=\frac{4\hbar ^{2}k_{\mathrm{o}}^{4}\left( 2k_{\mathrm{o}%
}^{2}+\sigma _{k\mathrm{o}}^{2}\right) R_{\mathrm{o}}}{3\mu },
\label{Vdensity}
\end{equation}%
which indicates the uniform scattering potential density is solely
determined by the initial conditions of the given system.

From (\ref{eq:p_0}), (\ref{eq:p_r}) and (\ref{eq:p_compare}) it is observed
that for the micro-correlated Gaussian system considered here, more time is
required to attain the same momentum value compared with the non-correlated
Gaussian system. For example, in order to attain the same value as the
initial momentum $p_{\mathrm{o}}$, the non-correlated system and the
micro-correlated system would require time intervals $\tau _{\mathrm{o}}$
and $\tau _{\ast }$, respectively, where 
\begin{eqnarray}
p_{\mathrm{o}} &=&\sqrt{p_{\mathrm{o}}^{2}+2\sigma _{\mathrm{o}}^{2}}\tanh
\left( A_{\mathrm{o}}\tau _{\mathrm{o}}\right) ,  \label{eq:p0_tau0} \\
p_{\mathrm{o}} &=&\sqrt{\left( 1-r\right) \left( p_{\mathrm{o}}^{2}+2\sigma
_{\mathrm{o}}^{2}\right) }\tanh \left( A_{\mathrm{o}}\tau _{\ast }\right) .
\label{eq:p0_tau*}
\end{eqnarray}%
Combining (\ref{eq:p0_tau0}) and (\ref{eq:p0_tau*}), we obtain%
\begin{equation}
\tanh \left( A_{\mathrm{o}}\tau _{\ast }\right) =\left( 1-r\right)
^{-1/2}\tanh \left( A_{\mathrm{o}}\tau _{\mathrm{o}}\right) .
\label{eq: tanh1}
\end{equation}%
Rewriting and expanding both sides of (\ref{eq: tanh1}), we have%
\begin{equation}
1-2e^{-2A_{\mathrm{o}}\tau _{\star }}+\mathcal{O}\left( e^{-4A_{\mathrm{o}%
}\tau _{\star }}\right) =\left( 1-r\right) ^{-1/2}\left[ 1-2e^{-2A_{\mathrm{o%
}}\tau _{\mathrm{o}}}+\mathcal{O}\left( e^{-4A_{\mathrm{o}}\tau _{\mathrm{o}%
}}\right) \right] .  \label{eq:tanh2}
\end{equation}%
Rounding (\ref{eq:tanh2}) off and arranging terms we obtain%
\begin{equation}
e^{-2A_{\mathrm{o}}\left( \tau _{\star }-\tau _{\mathrm{o}}\right) }\approx
\left( 1-r\right) ^{-1/2}-\frac{1}{2}\left[ \left( 1-r\right) ^{-1/2}-1%
\right] e^{2A_{\mathrm{o}}\tau _{\mathrm{o}}}.  \label{eq:tanh4}
\end{equation}%
The first term on the right hand side of (\ref{eq:tanh4}) can be
approximated to $1$ since $\left( 1-r\right) ^{-1/2}=1+\frac{1}{2}r+\mathcal{%
O}\left( r^{2}\right) $ and $r\ll 1$. However, $r$ in the second term should
not be disregarded in the same manner because $\left[ \left( 1-r\right)
^{-1/2}-1\right] e^{2A_{\mathrm{o}}\tau _{\mathrm{o}}}=\left[ \frac{1}{2}r+%
\mathcal{O}\left( r^{2}\right) \right] e^{2A_{\mathrm{o}}\tau _{\mathrm{o}}}$
is not negligible. Therefore, we may rewrite (\ref{eq:tanh4}) as%
\begin{equation}
e^{-2A_{\mathrm{o}}\Delta }\approx 1-\left[ \left( 1-r\right) ^{-1/2}-1%
\right] \cdot \eta _{\Delta },  \label{eq:Delta_tau0}
\end{equation}%
where $\Delta \equiv \tau _{\ast }-\tau _{\mathrm{o}}$ represents a new
quantity that we term "prolongation", and 
\begin{equation}
\eta _{\Delta }\equiv \frac{1}{2}e^{2A_{\mathrm{o}}\tau _{\mathrm{o}%
}}=\left( \frac{p_{\mathrm{o}}}{\sigma _{\mathrm{o}}}\right) ^{2}\exp \left[
\left( \frac{\sigma _{\mathrm{o}}}{p_{\mathrm{o}}}\right) ^{2}-\frac{3}{4}%
\left( \frac{\sigma _{\mathrm{o}}}{p_{\mathrm{o}}}\right) ^{4}+\mathcal{O}%
\left[ \left( \frac{\sigma _{\mathrm{o}}}{p_{\mathrm{o}}}\right) ^{6}\right] %
\right]  \label{eta}
\end{equation}%
for $\frac{\sigma _{\mathrm{o}}}{p_{\mathrm{o}}}\ll 1$ due to (\ref{a0}).
The quantities $\tau _{\ast }$ and $\tau _{\mathrm{o}}$ are the temporal
intervals required for a particle to reach the same value of momentum $k_{%
\mathrm{o}}$ from $0$ in the post-collisional scenario, in presence and in
the absence of correlations $r$, respectively. From (\ref{eq:Delta_tau0}) we
find 
\begin{equation}
\Delta \left( k_{\mathrm{o}},\sigma _{\mathrm{o}},r\right) \propto
\left\vert \ln \left\{ 1-\left[ \left( 1-r\right) ^{-1/2}-1\right] \cdot
\eta _{\Delta }\right\} \right\vert .  \label{duration}
\end{equation}%
Here, we can find the upper bound value of $r$ by means of (\ref{duration})
and (\ref{eta}), 
\begin{equation}
r<\frac{2}{\eta _{\Delta }}.  \label{r_bound}
\end{equation}%
The prolongation serves to quantify the time required by a micro-correlated
system - relative to a corresponding non-correlated one - to attain the same
momentum value (relative to the same initial reference time). The occurrence
of a non-vanishing prolongation is in fact due to the existence of
micro-correlations and therefore, due to the existence of scattering phase
shifts. In other words, in the absence of scattering there is no time
difference. This can be stated in yet another way as follows:
\textquotedblleft The prolongation encodes information about how long it
would take an entangled system to overcome the momentum gap (relative to a
corresponding non-entangled system) generated by the scattering phase shift.
The entangled system only attains the full value of momentum (i.e. the
momentum value as seen in the corresponding non-entangled system) when the
scattering phase shift vanishes. For this reason, the prolongation
represents the temporal duration over which the entanglement is
active\textquotedblright . We observe that the entanglement duration can be
controlled via the initial parameters $k_{\mathrm{o}}$, $\sigma _{\mathrm{o}%
} $ and the correlations $r$ (therefore via the incident particle energies
and the scattering potential due to (\ref{eq:corr_r1})). Moreover, we notice
that in the absence of correlations (i.e. $r\rightarrow 0$), $\Delta
\rightarrow 0$. It is anticipated that the maximum duration would be
obtained when $r$ is the greatest and the ratio $\sigma _{\mathrm{o}}/k_{%
\mathrm{o}}$ is the smallest.

The line element $ds^{2}=g_{ab}\left( \Theta \right) d\vartheta
^{a}d\vartheta ^{b}$ ($a,b=1,2,3$) on $\mathcal{M}_{s}^{\left( \text{corr.}%
\right) }$ is given by%
\begin{equation}
ds_{\mathcal{M}_{s}^{\left( \text{corr.}\right) }}^{2}=\frac{1}{\sigma ^{2}}%
\left( \frac{1}{1-r^{2}}d\mu _{x}^{2}+\frac{1}{1-r^{2}}d\mu _{y}^{2}-\frac{2r%
}{1-r^{2}}d\mu _{x}d\mu _{y}+4d\sigma ^{2}\right) ,
\end{equation}%
where we consider positive micro-correlation coefficients $r\in \left(
0,1\right) $. In this geometry, the components of the sectional curvature
are given by%
\begin{equation}
\mathcal{K}_{\mu _{1}}=-\frac{1}{4}=\mathcal{K}_{-\mu _{1}},\text{ }\mathcal{%
K}_{\mu _{2}}=-\frac{1}{4}=\mathcal{K}_{-\mu _{2}},\text{ }\mathcal{K}%
_{\sigma }=-\frac{1}{4}=\mathcal{K}_{-\sigma }.
\end{equation}%
The Ricci scalar $\mathcal{R}_{\mathcal{M}_{s}^{\left( \text{corr.}\right)
}} $ and Weyl Projective $\mathcal{W}_{abcd}$ curvatures are given by%
\begin{equation}
\mathcal{R}_{\mathcal{M}_{s}^{\left( \text{corr.}\right) }}=-\frac{3}{2}=%
\mathcal{R}_{\mathcal{M}_{s}^{\left( \text{uncorr.}\right) }}\text{ and }%
\mathcal{W}_{abcd}=0,  \label{scalar}
\end{equation}%
respectively. The fact that\textbf{\ }$\mathcal{W}_{abcd}=0$\textbf{\ }%
implies the manifold $\mathcal{M}_{s}^{\left( \text{corr.}\right) }$\textbf{%
\ }is isotropic.\textbf{\ }It is known that the anisotropy of the manifold
underlying system dynamics plays a crucial role in the mechanism of
instability. In particular, fluctuating sectional curvatures require also
that the manifold be anisotropic.

It can be shown \cite{PLA} that the JLC-equation (\ref{JLC-1}) on $\mathcal{M%
}_{s}^{\left( \text{corr.}\right) }$ reduces to%
\begin{equation}
\frac{D^{2}\mathcal{J}_{\mathcal{M}_{s}^{\left( \text{corr.}\right) }}}{%
D\tau ^{2}}+Q\mathcal{J}_{\mathcal{M}_{s}^{\left( \text{corr.}\right) }}=0.
\label{eq:VII-11}
\end{equation}%
where $Q\equiv \frac{\mathcal{R}\left\Vert \mathbf{v}\right\Vert ^{2}}{n(n-1)%
}=-A_{_{\mathrm{o}}}^{2}<0$ and $\mathcal{J}_{\mathcal{M}_{s}^{\left( \text{%
corr.}\right) }}$ is defined in (\ref{eq:VII-9}). Since $Q<0$, unstable
solutions of equation (\ref{eq:VII-11}) assumes the form%
\begin{equation}
\mathcal{J}_{\mathcal{M}_{s}^{\left( \text{corr.}\right) }}\left( \tau
\right) =\frac{1}{\sqrt{-Q}}\omega \left( 0\right) \sinh \left( \sqrt{-Q}%
\tau \right) ,  \label{jac}
\end{equation}%
where $\omega \left( 0\right) \equiv \left. \frac{d\mathcal{J}_{\mathcal{M}%
_{s}^{\left( \text{corr.}\right) }}\left( \tau \right) }{d\tau }\right\vert
_{\tau =0}$. Recalling the definition of the hyperbolic sine function $\sinh
x=\frac{1}{2}\left( e^{x}-e^{-x}\right) $, it is clear that the geodesic
deviation on $\mathcal{M}_{s}^{\left( \text{corr.}\right) }$ is described by
means of an exponentially divergent Jacobi vector field intensity $\mathcal{J%
}_{\mathcal{M}_{s}^{\left( \text{corr.}\right) }}$, a classical feature of
chaos. In order to evaluate (\ref{lap}) we use (\ref{jac}) to find $%
\left\vert \mathcal{J}_{\mathcal{M}_{s}^{\left( \text{corr.}\right) }}\left(
\tau \right) \right\vert ^{2}=\frac{\omega ^{2}\left( 0\right) }{-Q}\sinh
^{2}\left( \sqrt{-Q}\tau \right) $ and $\left\vert \frac{d\mathcal{J}_{%
\mathcal{M}_{s}^{\left( \text{corr.}\right) }}\left( \tau \right) }{d\tau }%
\right\vert ^{2}=\omega ^{2}\left( 0\right) \cosh ^{2}\left( \sqrt{-Q}\tau
\right) $. Thus, for the case being considered, the Lyapunov exponents $%
\lambda _{\mathcal{M}_{s}^{\left( \text{corr.}\right) }}=\lim_{\tau
\rightarrow \infty }\frac{1}{\tau }\ln \left[ \frac{1}{4}\left( 1-Q\right)
e^{2\sqrt{-Q}\tau }\right] =2\sqrt{-Q}.$ Therefore, it follows that%
\begin{equation}
\lambda _{\mathcal{M}_{s}^{\left( \text{corr.}\right) }}\overset{\tau
\rightarrow \infty }{=}2\sqrt{-Q}=2A_{\mathrm{o}}>0.  \label{due}
\end{equation}%
From (\ref{due}) we observe the following points: the classical chaoticity
does not depend on the statistical correlation, i.e. $\lambda _{\mathcal{M}%
_{s}^{\left( \text{corr.}\right) }}=\lambda _{\mathcal{M}_{s}^{\left( \text{%
non-corr.}\right) }}\equiv \lambda _{\mathcal{M}_{s}}=2A_{\mathrm{o}}$, and
the\textbf{\ }Lyapunov exponents can be determined solely from the initial
conditions (see equation (\ref{a0})).

Yet another finding uncovers an interesting quantitative connection between
quantum entanglement quantified by the purity $\mathcal{P}$ in (\ref%
{eq:purity}) and the information geometric complexity (IGC) of motion on the
uncorrelated and correlated curved statistical manifolds $\mathcal{M}%
_{s}^{\left( \text{uncorr.}\right) }$ and $\mathcal{M}_{s}^{\left( \text{%
corr.}\right) }$, respectively. The information geometric complexity as
defined in \cite{cafaroAMC} represents the volume of the effective
parametric space explored by the system in its evolution between the chosen
initial and final macrostates. The volume itself is in general given in
terms of a multidimensional integral over the geodesic paths connecting the
initial and final macrostates. For additional details see \cite{cafaroAMC}.
Here, omitting technical details and following the works presented in \cite%
{PLA, cafaroPD, cafaroPA2010}, one finds%
\begin{equation}
\mathcal{C}_{\text{IG}}^{\left( \text{corr.}\right) }\left( \tau ;r\right) =%
\frac{8\sqrt{\frac{1-r}{1+r}}}{\lambda _{\mathcal{M}_{s}}}\left[ -\frac{3}{4}%
\lambda _{\mathcal{M}_{s}}+\frac{1}{4}\frac{\sinh \left( \lambda _{\mathcal{M%
}_{s}}\tau \right) }{\tau }+\frac{\tanh \left( \frac{1}{2}\lambda _{\mathcal{%
M}_{s}}\tau \right) }{\tau }\right] ,  \label{complexity}
\end{equation}%
where $\mathcal{C}_{\text{IG}}^{\left( \text{corr.}\right) }$ denotes the
IGC on $\mathcal{M}_{s}^{\left( \text{corr.}\right) }$. Similarly, $\mathcal{%
C}_{\text{IG}}^{\left( \text{uncorr.}\right) }=\mathcal{C}_{\text{IG}%
}^{\left( \text{corr.}\right) }\left( r\rightarrow 0\right) $ represents the
IGC on $\mathcal{M}_{s}^{\left( \text{uncorr.}\right) }$. As a side remark,
we point out that (\ref{complexity}) confirms that an increase in the
correlational structure of the dynamical equations for the statistical
variables labelling a macrostate of a system implies a reduction in the
complexity of the geodesic paths on the underlying curved statistical
manifolds \cite{CM, cafaroPD}. In other words, making macroscopic
predictions in the presence of correlations is easier than in their absence.

The technical details that will be omitted in what follows may be found in 
\cite{CM}. By direct computation, the IGE is found to be%
\begin{equation}
\mathcal{S}_{\mathcal{M}_{s}^{\left( \text{corr.}\right) }}\left( \tau
;r\right) \overset{\tau \rightarrow \infty }{=}\lambda _{\mathcal{M}%
_{s}}\tau -\ln \left( \lambda _{\mathcal{M}_{s}}\tau \right) +\frac{1}{2}\ln
\left( \frac{1-r}{1+r}\right) .  \label{eq:entrp_r}
\end{equation}%
For non-correlated Gaussian statistical models the IGE is given by $\mathcal{%
S}_{\mathcal{M}_{s}^{\left( \text{uncorr.}\right) }}\left( \tau ;0\right) =%
\mathcal{S}_{\mathcal{M}_{s}^{\left( \text{corr.}\right) }}\left( \tau
;r\rightarrow 0\right) $. Observe that in contrast to the macro-correlated
case, the IGE for the micro-correlated case presents linear growth in the
affine temporal parameter $\tau $. Both the IGC and the IGE decrease in
presence of micro-correlations.\ In particular, the IGC decreases by the
factor $\sqrt{\frac{1-r}{1+r}}<1$ for $r>0$ whereas\textbf{\ }the IGE
decreases by $\frac{1}{2}\ln \left( \frac{1-r}{1+r}\right) <0$ for $r>0$.
With the quantities $\mathcal{C}_{\text{IG}}^{\left( \text{corr.}\right)
}\left( \tau \text{; }r\right) $ and $\mathcal{S}_{\mathcal{M}_{s}^{\left( 
\text{corr.}\right) }}\left( \tau \text{; }r\right) $ in hand, we make the
following interesting observations. From (\ref{complexity}) we find%
\begin{equation}
r=\frac{\Delta \mathcal{C}^{2}}{\mathcal{C}_{\text{total}}^{2}},
\label{due2}
\end{equation}%
where 
\begin{equation}
\Delta \mathcal{C}^{2}\equiv \left[ \mathcal{C}_{\text{IG}}^{\left( \text{%
uncorr.}\right) }\right] ^{2}-\left[ \mathcal{C}_{\text{IG}}^{\left( \text{%
corr.}\right) }\right] ^{2}
\end{equation}%
and%
\begin{equation}
\mathcal{C}_{\text{total}}^{2}\equiv \left[ \mathcal{C}_{\text{IG}}^{\left( 
\text{uncorr.}\right) }\right] ^{2}+\left[ \mathcal{C}_{\text{IG}}^{\left( 
\text{corr.}\right) }\right] ^{2}.
\end{equation}%
Combining (\ref{eq:purity}) and (\ref{complexity}) it follows that%
\begin{equation}
\mathcal{P}\approx 1-\eta _{\mathcal{C}}\cdot \frac{\Delta \mathcal{C}^{2}}{%
\mathcal{C}_{\text{total}}^{2}},  \label{ultima}
\end{equation}%
where the dimensionless coefficient $\eta _{\mathcal{C}}$ reads%
\begin{equation}
\eta _{\mathcal{C}}\equiv \frac{8}{3}k_{\mathrm{o}}^{2}\left( 2k_{\mathrm{o}%
}^{2}+\sigma _{k_{\mathrm{o}}}^{2}\right) R_{\mathrm{o}}L^{3}.
\end{equation}%
From (\ref{ultima}) it is evident that the scattering-induced quantum
entanglement and the information geometric complexity of motion are
connected. In particular, when purity approaches to unity (entanglement-free
scenario), the difference between the correlated and uncorrelated
information geometric complexities approaches zero.

\section{Conclusions}

In this article, we have introduced a theoretical construct that allows us
to describe the macroscopic behavior of complex systems in terms of the
underlying statistical structure of their microscopic degrees of freedom
through statistical inference and information geometry methods. We reviewed
the Maximum relative Entropy (MrE) formalism and the theoretical structure
of the information geometrodynamical approach to chaos (IGAC) on curved
statistical manifolds $\mathcal{M}_{S}$. Special focus was devoted to the
description of the roles played by the sectional curvature $\mathcal{K}_{%
\mathcal{M}_{S}}$, the Jacobi field intensity $\mathcal{J}_{\mathcal{M}_{S}}$
and the information geometrodynamical entropy $\mathcal{S}_{\mathcal{M}_{S}}$
(IGE) as information geometric indicators of chaoticity (complexity). Four
applications of these information geometric techniques combined with ME
methods were presented.

First, we studied the chaotic behavior of a Gaussian statistical model
describing an arbitrary system of $l$ uncorrelated degrees of freedom and
found that the hyperbolicity of the non-maximally symmetric $2l$-dimensional
statistical manifold $\mathcal{M}_{s}$ underlying such a Gaussian model
leads to linear IGE growth and to exponential divergence of the Jacobi
vector field intensity \cite{ca-6}. Second, we studied the asymptotic
behavior of the dynamical complexity of the maximum probability trajectories
on Gaussian statistical manifolds in presence of correlation-like terms
between macrovariables labeling the macrostates of the system under
investigation. In presence of correlation-like terms, we observed a power
law decay of the information geometric complexity at a rate determined by
the correlation coefficient. We also presented an information-geometric
analogue of the Zurek-Paz quantum chaos criterion of linear entropy growth.
This analogy was motivated by studying the information geometrodynamics of
an ensemble of random frequency macroscopic inverted harmonic oscillators.
The IGAC was also employed to study the entropic dynamics on curved
statistical manifolds induced by classical probability distributions
commonly used in the study of regular and chaotic quantum energy level
statistics. In doing so, we suggest an information-geometric
characterization of regular and chaotic quantum energy level statistics.

Finally, the IGAC was used to describe the scattering-induced quantum
entanglement between two Gaussian wave-packets. The IGAC was used to analyze
our specific two-variable micro-correlated Gaussian statistical model. The
manifolds $\mathcal{M}_{s}^{\left( \text{corr.}\right) }$ and $\mathcal{M}%
_{s}^{\left( \text{uncorr.}\right) }$ were used to model the quantum
entanglement induced by head-on scattering (in the $s$-wave approximation)
of two spinless, structureless, non-relativistic particles, each represented
by minimum uncertainty wave-packets. Equation (\ref{eq:phase_shift_V})
allowed to connect the entanglement strength - quantified in terms of purity
- to the scattering potential and incident particle energies ((\ref%
{eq:corr_r1}) and (\ref{eq:purity})). It was also found to be possible to
relate the statistical entanglement duration $\Delta $ to the scattering
potential $V\left( x\right) $\ and incident particle energies $\mathcal{E}$\
((\ref{eq:corr_r1}) and (\ref{duration})). Recall that the prolongation $%
\Delta $ was defined as the time required for the observed momentum
difference between a correlated and corresponding non-correlated system to
vanish. The prolongation encodes information about how long it would take an
entangled system to overcome the momentum gap generated by the scattering
phase shift. The entangled system only attains the full value of momentum%
\textbf{\ }(i.e. the momentum value as seen in the corresponding
non-correlated system) when the scattering phase shift vanishes.\textbf{\ }%
For this reason, the prolongation represents the temporal duration over
which the entanglement is active.

The micro-correlation coefficient $r$, a quantity that parameterizes the
correlated microscopic degrees of freedom of the system, can be understood
as the ratio of the potential to kinetic energy of the system. When $r\neq 0$
the wave-packets experience the effect of a repulsive potential; the
magnitude of the wave vectors (momenta) decreases relative to their
corresponding non-correlated value. The upper bound value of $r$ depends on $%
p_{\mathrm{o}}$ and $\sigma _{\mathrm{o}}$ in such a manner that $r$
increases as $p_{\mathrm{o}}$ decreases. This result constitutes a
significant, explicit connection between micro-correlations (the correlation
coefficient $r$) and physical observables (the macrovariable $p_{\mathrm{o}}$%
). For $r$ values close to its upper bound, the prolongation $\Delta $
becomes infinitely large. On the other hand, with $r$ vanishing (i.e., no
micro-correlation) $\Delta $ is identically zero. With $r$ fixed however,
the prolongation $\Delta $ depends on $p_{\mathrm{o}}$ and $\sigma _{\mathrm{%
o}}$. Thus, the prolongation $\Delta $ can be controlled by the initial
conditions $p_{\mathrm{o}}$ and $\sigma _{\mathrm{o}}$ as well as $r$.
Maximal prolongation occurs when $r$ is greatest and the ratio $\sigma _{%
\mathrm{o}}/p_{\mathrm{o}}$ is smallest. For small initial $r$ and $p_{%
\mathrm{o}}$, $\Delta $ would be correspondingly small, suggesting that for
such scenarios quantum entanglement is transient. Furthermore, a
quantitative relation between quantum entanglement (purity) and the IGC (\ref%
{ultima}) was uncovered.

The complexity of geodesic paths on $\mathcal{M}_{s}^{\left( \text{corr.}%
\right) }$ and $\mathcal{M}_{s}^{\left( \text{uncorr.}\right) }$ was
characterized through the asymptotic computation of the IGE and the\textbf{\ 
}Lyapunov exponents on each manifold. The Lyapunov exponents in both cases
were found to be the same positive definite constant, $\lambda _{\mathcal{M}%
_{s}^{\left( \text{corr.}\right) }}=\lambda _{\mathcal{M}_{s}^{\left( \text{%
uncorr.}\right) }}\equiv \lambda _{\mathcal{M}_{s}}=2A_{\mathrm{o}}>0$. The
IGE $\mathcal{S}_{\mathcal{M}_{s}^{\left( \text{corr.}\right) }}\left( \tau
;r\right) $ in presence of micro-correlations assumes a smaller initial
value relative to the non-correlated case $\mathcal{S}_{\mathcal{M}%
_{s}^{\left( \text{uncorr.}\right) }}\left( \tau ;0\right) $ while the
growth characteristics of both correlated and non-correlated IGEs were found
to be similar. Specifically, the larger the micro-correlation (i.e. the
closer $r$ is to $1$) the lower the initial value of the IGE. Thus, the
stronger the initial micro-correlation, the larger the gap between $\mathcal{%
S}_{\mathcal{M}_{s}^{\left( \text{corr.}\right) }}\left( \tau =0;r\right) $\
and $\mathcal{S}_{\mathcal{M}_{s}^{\left( \text{uncorr.}\right) }}\left(
\tau =0;0\right) $. This implies that $\mathcal{S}_{\mathcal{M}_{s}^{\left( 
\text{corr.}\right) }}\left( \tau ;r\right) <\mathcal{S}_{\mathcal{M}%
_{s}^{\left( \text{uncorr.}\right) }}\left( \tau ;0\right) $. When
micro-correlations vanish (i.e. when $r=0$), we obtain the expected result $%
\mathcal{S}_{\mathcal{M}_{s}^{\left( \text{corr.}\right) }}\left( \tau
;0\right) =\mathcal{S}_{\mathcal{M}_{s}^{\left( \text{uncorr.}\right)
}}\left( \tau \text{; }0\right) $. The appearance of micro-correlation terms
in the elements in the Fisher-Rao information metric leads to the
compression of $\mathcal{C}_{\text{IG}}^{\left( \text{corr.}\right) }\left(
\tau ;r\right) $ by the fraction $\sqrt{\frac{1-r}{1+r}}$ and thus, to a
reduction of the complexity of the path leading from $\Theta ^{\text{%
(initial)}}$ to $\Theta ^{\text{(final)}}$.

We emphasize that at this stage of development, IGAC remains an ambitious
unifying\ information-geometric theoretical construct for the study of
chaotic dynamics with several unsolved problems. However, based on our
findings, we believe it provides an interesting, innovative and potentially
powerful way to study and understand the very important and challenging
problems of classical and quantum chaos through statistical inference and
information geometric techniques. In order to clarify the relationship
between our IGE and conventional measures of complexity, for instance the
topological entropy (the supremum of the Kolmogorov-Sinai metric entropy),
further investigation is required. Finally, we believe our IGE may play an
important role in both classical and quantum information science, but at
this moment this remains a conjecture \cite{ca-5, ca-7}.

\emph{One final remark}: In this work, we have used information geometric
techniques and inductive inference methods for tackling some computational
problems of interest in classical and quantum physics. Specifically, we have
provided an information geometric characterization of the complexity of
dynamical systems in terms of their probabilistic description on curved
statistical manifolds. The maximally probable trajectories of the system
arise through implementation of a Principle of Inference, the ME method and
the constructed indicators of complexity of such trajectories are defined in
terms of asymptotic temporal averages. This leads to the following
consideration. Most standard characterizations of chaoticity are based on
the Boltzmann's ergodic hypothesis which relies on a \textquotedblleft
frequency\textquotedblright\ interpretation of probabilities where concepts
such as coarse graining (or, randomization) appear. Such artificial concepts
are not needed in our information geometric characterization based on ME
methods \cite{adom-tesi} and they may even lead to incorrect results \cite%
{jay1, jay2}. Does this mean that ergodic theory is unnecessary for our
purposes \cite{jay3}? Certainly, ergodic theory is explicitly used in the
characterization of chaos in terms of the Kolmogorov-Sinai dynamical entropy
and Lyapunov exponents \cite{Eck}. It appears that it is implicitly used in
the formulation of the IGAC and the IGE by way of focusing on the above
mentioned asymptotic temporal averages. Thus it remains to be seen whether
or not the IGAC and the IGE are \textquotedblleft special
cases\textquotedblright\ of a broader picture.

\begin{acknowledgments}
C. C. thanks Ariel Caticha, John Kimball, Kevin Knuth, Stefano Mancini and
Carlos Rodriguez for useful discussions and/or comments on the application
of information geometry and inference methods to chaos. D.H.K. acknowledges
the support of the National Research Foundation of Korea (NRF) grant funded
by the Korea government (MEST) through the Center for Quantum Spacetime
(CQUeST) of Sogang University with grant number 2005-0049409. D.H.K. also
acknowledges the support of the World Class University (WCU) program of
NRF/MEST (R32-2009-000-10130-0).
\end{acknowledgments}

\end{document}